\pdfoutput=1
\RequirePackage{color}
\RequirePackage{ifpdf}
\documentclass{JINST}

\usepackage{graphicx}
\usepackage{multirow}
\usepackage{subfig}
\usepackage{lineno}
\usepackage{tablefootnote}
\makeatletter
\AtBeginDocument{%
\DeclareRobustCommand*\subref{\@ifstar\sf@@subref\sf@subref}}
\makeatother
\usepackage{url}
\makeatletter
\def\url@leostyle{%
  \@ifundefined{selectfont}{\def\UrlFont{\sf}}{\def\UrlFont{\small\ttfamily}}}
\makeatother
\title{The Electronics, Trigger and Data Acquisition System for the Liquid Argon Time Projection Chamber of the DarkSide-50 Search for Dark Matter}

\author{
P.~Agnes$^{af,s}$,
I.~F.~M.~Albuquerque$^{ap}$,
T.~Alexander$^{ae}$,
A.~K.~Alton$^{b}$,
K.~Arisaka$^{at}$,
D.~M.~Asner$^{ae}$,
M.~Ave$^{ap}$,
H.~O.~Back$^{aj,ae}$,
B.~Baldin$^{m}$,
K.~Biery$^{m}$
% ...
\\
 (DarkSide Collaboration)
\\
\llap{$^{a}$}Amherst Center for Fundamental Interactions and Dept. of Physics, University of Massachusetts, Amherst, MA 01003, USA\\
\llap{$^{b}$}Dept. of Physics, Augustana University, Sioux Falls, SD 57197, USA \\
\llap{$^{c}$}Radiation Physics Laboratory, Belgorod National Research University, Belgorod 308007, Russia \\
\llap{$^{d}$}Inst. for High Energy Physics, Beijing 100049, China\\
\llap{$^{e}$}School of Physics, University of Chinese Academy of Sciences, Beijing 100049, China\\
\llap{$^{f}$}School of Natural Sciences, Black Hills State University, Spearfish, SD 57799, USA \\
\llap{$^{g}$}Dept. of Physics, Universit\`a degli Studi, Cagliari 09042, Italy \\
\llap{$^{h}$}INFN, Sezione di Cagliari, Cagliari 09042, Italy  \\
\llap{$^{i}$}Laboratorio Subterr\'aneo de Canfranc, Canfranc Estaci\'on E-22880, Spain  \\
\llap{$^{j}$}Kavli Inst., E.F.I. and Dept. of Physics, University of Chicago, Chicago, IL 60637, USA \\
\llap{$^{k}$}Dept. of Physics, University of California, Davis, CA 95616, USA \\
\llap{$^{l}$}Joint Inst. for Nucl. Research, Dubna 141980, Russia \\
\llap{$^{m}$}Fermi National Accelerator Laboratory, Batavia, IL 60510, USA \\
\llap{$^{n}$}St. Petersburg Nucl. Physics Inst. NRC Kurchatov Inst., Gatchina 188350, Russia \\
\llap{$^{o}$}Dept. of Physics, Universit\`a degli Studi, Genova 16146, Italy \\
\llap{$^{p}$}INFN, Sezione di Genova, Genova 16146, Italy\\
\llap{$^{q}$}Gran Sasso Science Inst., L'Aquila 67100, Italy \\
\llap{$^{r}$}Dept. of Physics and Astronomy, University of Hawai'i, Honolulu, HI 96822, HI \\
\llap{$^{s}$}Dept. of Physics, University of Houston, Houston, TX 77204, USA \\
\llap{$^{t}$}Inst. for National Research, National Academy of Sciences of Ukraine, Kiev 03680, Ukraine\\
\llap{$^{u}$}Smoluchowski Inst. of Physics, Jagiellonian University, Krakow 30059, Poland \\
\llap{$^{v}$}Lawrence Livermore National Laboratory, Livermore, CA 94550, USA \\
\llap{$^{w}$}Laboratori Nazionali del Gran Sasso, Assergi AQ 67010, Italy \\
\llap{$^{x}$}Dept. of Physics, Universit\`a degli Studi, Milano 20133, Italy \\
\llap{$^{y}$}INFN, Sezione di Milano, Milano 20133, Italy\\
\llap{$^{z}$}National Research Centre Kurchatov Institute, Moscow 123182, Russia \\
\llap{$^{aa}$}National Research Nucl. University, MEPhI, Moscow 115409, Russia \\
\llap{$^{ab}$}Skobeltsyn Inst. of Nucl. Physics, Lomonosov Moscow State University, Moscow 119991, Russia \\
\llap{$^{ac}$}Dept. of Physics, Universit\`a degli Studi Federico II, Napoli 80126, Italy \\
\llap{$^{ad}$}INFN, Sezione di Napoli, Napoli 80126, Italy \\
\llap{$^{ae}$}Pacific Northwest National Laboratory, Richland, WA 99354, USA \\
\llap{$^{af}$}APC, Universit\'e Paris Diderot, CNRS/IN2P3, CEA/Irfu, Obs de Paris, USPC, 75205 Paris, France \\
\llap{$^{ag}$}LPNHE Paris, Universit\'e Pierre et Marie Curie,  CNRS/IN2P3, Paris 75252, France \\
\llap{$^{ah}$}Dept. of Chemistry, Biology and Biotechnology, Universit\`a degli Studi, Perugia 06123, Italy \\
\llap{$^{ai}$}INFN, Sezione di Perugia, Perugia 06123, Italy \\
\llap{$^{aj}$}Dept. of Physics, Princeton University, Princeton, NJ 08544, USA\\
\llap{$^{ak}$}Physics Dept., Sapienza Universit\'a di Roma, Roma 00185, Italy \\
\llap{$^{al}$}INFN, Sezione di Roma, Roma 00185, Italy \\
\llap{$^{am}$}Dept. of Physics and Mathematics, Universit\`a degli Studi Roma Tre, Roma 00146, Italy \\
\llap{$^{an}$}INFN, Sezione di Roma Tre, Roma 00146, Italy\\
\llap{$^{ao}$}Federal University of ABC (UFABC), Av. dos Estados, 5001, Santo Andr\`e, SP, 09210-170, Brazil\\
\llap{$^{ap}$}Inst. de F\'isica, Universidade de S\~ao Paulo, S\~ao Paulo 05508-090, Brazil \\
\llap{$^{aq}$}Inst. of Physics Gleb Wataghin,  Universidade Estadual de Campinas, S\~ao Paulo 13083-859, Brazil\\
\llap{$^{ar}$}SLAC National Accelerator Laboratory, Menlo Park, CA 94025, USA \\
\llap{$^{as}$}Dept. of Physics, Temple University, Philadelphia, PA 19122, USA \\
\llap{$^{at}$}Dept. of Physics and Astronomy, University of California, Los Angeles, CA 90095, USA\\
\llap{$^{au}$}Dept. of Physics, Virginia Tech, Blacksburg, VA 24061, USA\\
}

\abstract{The DarkSide-50 experiment at the Laboratori Nazionali del Gran Sasso is a search for dark matter using a dual phase time projection chamber with 50 kg of low radioactivity argon as target.  Light signals from interactions in the argon are detected by a system of 38 photo-multiplier tubes (PMTs), 19 above and 19 below the TPC volume inside the argon cryostat.  We describe the electronics which processes the signals from the photo-multipliers, the trigger system which identifies events of interest, and the data-acquisition system which records the data for further analysis. The electronics include resistive voltage dividers on the PMTs, custom pre-amplifiers mounted directly on the PMT voltage dividers in the liquid argon, and custom amplifier/discriminators (at room temperature). After amplification, the PMT signals are digitized in CAEN waveform digitizers, and CAEN logic modules are used to construct the trigger; the data acquisition system for the TPC is based on the Fermilab {\it{artdaq}} software. The system has been in operation since early 2014.}


\begin{document}

\section{Introduction - The DarkSide-50 Detector}
\label{sec:Introduction}

\begin{figure}[h]
\begin{center}
\includegraphics[width=4.0in]{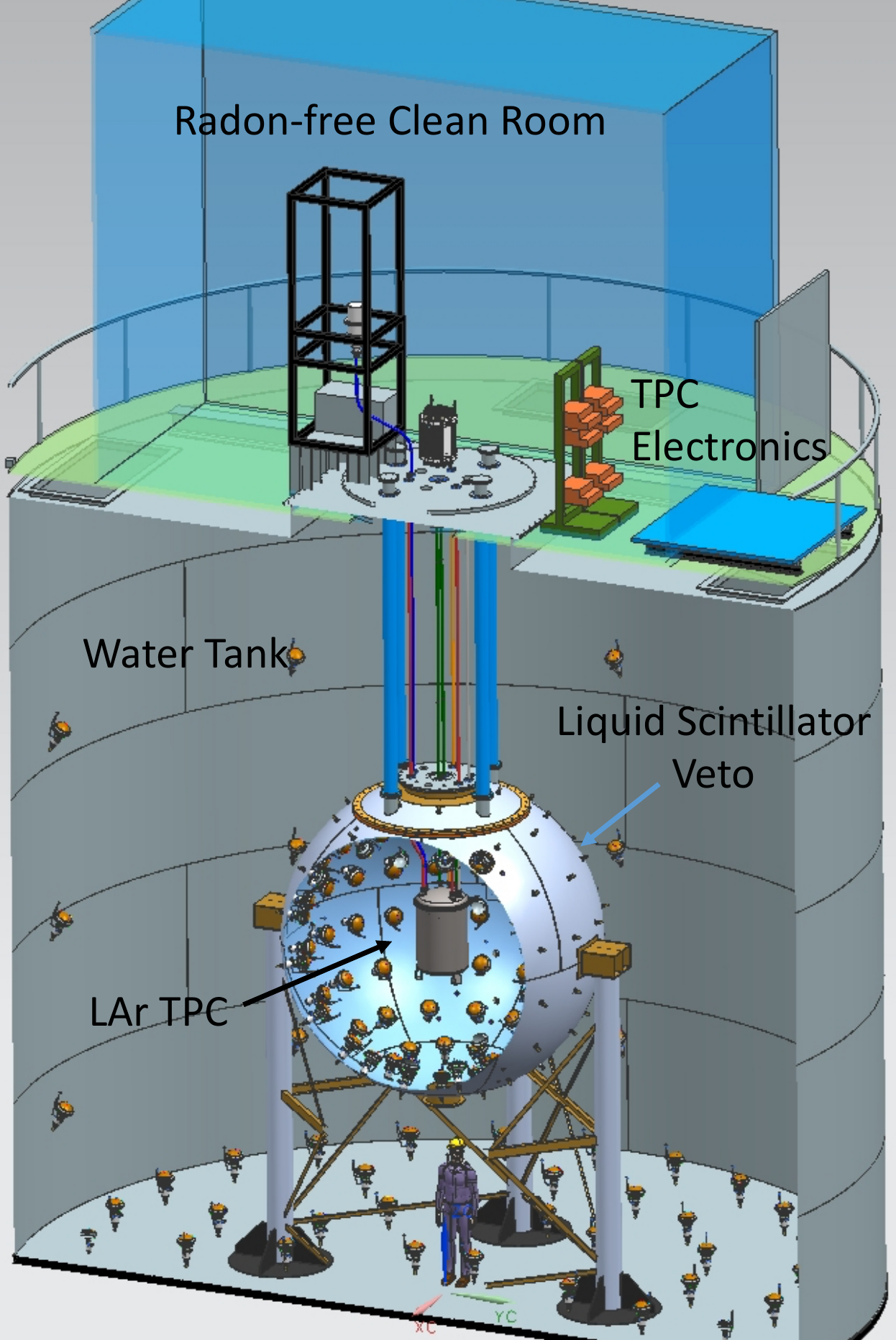}
\caption{A schematic of the DarkSide-50 Detector.}
\label{fig:DS-50-schematic}
\end{center}
\end{figure}

The DarkSide-50 experiment is a search for dark matter in the form of weakly interacting massive particles. The apparatus is located in the Laboratori Nazionali del Gran Sasso (LNGS) of the Italian Istituto Nazionale di Fisica Nucleare (INFN). At the heart of the detector is a two-phase argon time projection chamber (TPC) with an active mass of 50 kg (155 kg argon total). The TPC cryostat is suspended inside a 4 m diameter steel vessel filled with borated liquid scintillator that acts as a neutron veto. The neutron veto, in turn, is mounted inside a cylindrical water tank, 10 m high and 11 m in diameter, that serves to identify muons and acts as a shield against photons from the surrounding rock. Figure \ref{fig:DS-50-schematic} shows a schematic of the full apparatus. Results from a 50 day run (1440 kg-days) with atmospheric argon have been been published in \cite{ref:50days} and from a 70 day exposure (2600 kg-days) with low radioactivity argon in \cite{ref:70days}.  A description of the Veto system  and its electronics appear in  \cite{ref:Vetopaper} and \cite{ref:Vetoelectronics} respectively.

The active region of the two-phase argon TPC is a right cylinder of diameter 35.6 cm and height 37.1 cm formed by a hollow cylinder of low-radioactivity teflon capped with silica windows at the top and bottom.  Liquid is present in this volume to a height of 36.1 cm above which there is a 1 cm gas region.  Three electrodes,  the cathode at the bottom of the active volume, a 95\% transparent grid just below the liquid surface, and the anode at the top of the TPC gas region define electric potentials at the bottom of the liquid, near the liquid surface and at the top of the gas region. The cathode and anode are formed by thin layers of Indium Tin Oxide (ITO) on the top of the bottom and the bottom of the top silica window respectively. A uniform field region in the liquid between the cathode and the grid is established with a set of copper field-rings connected by a resistor chain. A second and more intense field is defined between the grid and the anode. When energy is deposited in the argon, the argon emits scintillation light (the S1 signal) and a number of ion-electron pairs are produced. The ionized electrons which escape recombination are drifted by the electric field in the liquid towards the liquid surface from which they are extracted into the argon gas region by the field between the grid and the anode. As the electrons pass through the gas region, they induce further emission of light by electro-fluorescence in the gas (the S2 signal). The wave-length shifter Tetraphenyl-butadiene (TPB) is used to convert the argon light, which has a wavelength of 128 nm in vacuum, to the visible. The TPB is applied to all the inner surfaces of the active volume. 

\begin{center}
\begin{figure}[h]
\includegraphics[width=2.5in]{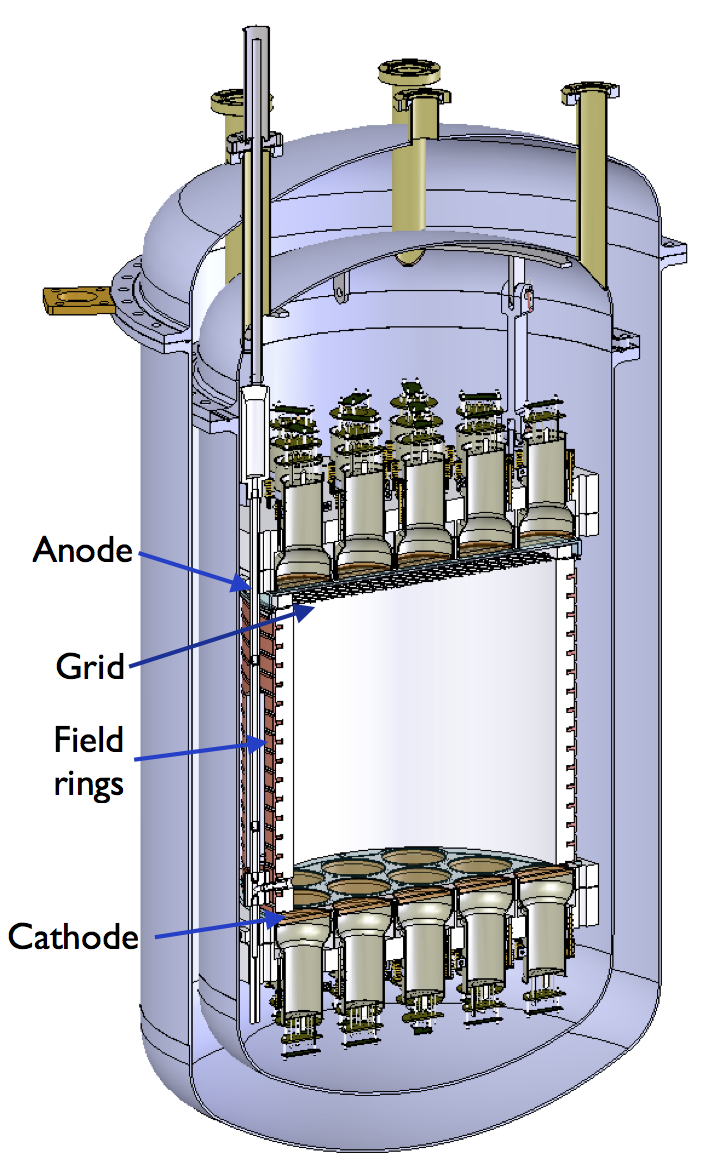} \hfill
\includegraphics[width=2.5in]{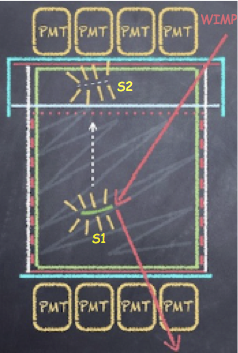}
\caption{Schematic of the DarkSide-50 TPC and its mode of operation. The spatial extent of the S1 and the S2 signals are an artistic license.}
\label{fig:TPC-schematic}
\end{figure}
\end{center}

The S1 and S2 signals are viewed through the silica windows by a system of 38 photo-multiplier tubes (PMTs) ~\cite{ref:Hamamatsu_PMT_model} , 19 immediately above the upper window and 19 immediately below the lower window.  (The top PMTs are also submerged in liquid to avoid the risk of sparking in the voltage dividers.)  The first signal, S1, gives information on the amount of energy deposited and, through its time profile, on the type of interaction in the liquid.  The second signal because of its proximity to the top PMTs gives information on the transverse position of the interaction and the time interval between S2 and S1 gives information on the depth within the liquid of the interaction given the known drift-velocity of electrons in the liquid argon. The original interaction can thus be localized in 3-D within a few mm. (A full description of the geometric reconstruction is given in \cite{ref:BrodskyThesis}.) Figure \ref{fig:TPC-schematic} shows a cut-away schematic of the TPC, and its mode of operation.

\section{Requirements on the Electronics for the TPC}
\label{sec:requirements}

Liquid argon is an attractive target medium for dark matter searches.  It is a bright scintillator, producing, at zero electric field,   $\sim$10\, photons per keV deposited by a recoiling argon nucleus  and $\sim$40\, photons per keV of energy deposited by electrons or photons  \cite{ref:ArLightYields}. A second and critical feature is that the time distribution of the scintillation can be used to distinguish between energy deposited in the liquid by an electron or photon, and energy transferred to the nucleus of an argon atom which then recoils in the liquid. (The latter would be the result of a dark matter interaction in the argon while the former is a major source of potential background.) 

The light in argon is produced by the decay of two excited dimer states with very different time constants; the  `singlet' excited state has a lifetime of $\sim$6 ns, while the  `triplet' excited state has a lifetime of $\sim$1.5\,$\mu$s in liquid.  Electron and nucleus recoils produce these states in different ratios, the singlet to triplet ratio being $\sim$1:2 for electron recoils while it is $\sim$3:1\, in the case of a recoiling nucleus. The resultant difference in the time profile of the scintillation allows the background from beta and gamma radioactivity in the materials of the detector to be distinguished from the signal of a recoiling argon nucleus \cite{ref:50days}. A third feature, relevant in the case of the dual-phase LAr TPC, is that the ratio of the S1 signal (prompt light) to the S2 signal (light from ionized electrons reaching the gas region of the TPC) is quite different for electron recoils (S2 typically >> S1) and nucleus recoils (S2 typically <S1), again providing discrimination between electromagnetic backgrounds and candidate events.  The actual ratios depend on the details of the TPC.  

The bandwidth required in the PMT, and the front-end electronics and digitizer, could depend on a number of parameters such as the singlet and triplet lifetimes, the decay time of the wave-length shifter, the range of the travel time of photons in the TPC before striking a PMT, and the drift speed of the ionized electrons. In practice the critical requirement is to be able to measure the singlet to triplet ratio to optimize the rejection of electromagnetic background.  Given the decay constants of the two states, a standard parameter used to characterize this ratio is the fraction of the total scintillation recorded in the first 90 ns, {\it f$_{90}$}. Our PMTs produce signals of $\sim$ 30 ns base to base (see figure \ref{fig:TPCPreamp-spe}) more than fast enough for the measurement of {\it f$_{90}$} and these signals are recorded with a bandwidth of 100 MHz. 

The sensitivity and dynamic range are set by the following considerations. The total dark matter interaction rate above a given energy threshold is predicted to be a falling exponential with a reduction of  $\sim2$ as the threshold increases by 20 keV \cite{ref:Snowmass2015}. At our operating electric field of 200 V/cm, our apparatus produces about two photo-electrons per keV for a recoil nucleus; hence the requirement that single photo-electrons be recorded with full efficiency. The dynamic range requirement is relatively soft.  For the dark matter search itself, there is not much to be gained by extending the search region beyond an energy transfer of $\sim100$ keV.  and so the  maximum number of photo-electrons in the fast component of the S1 signal of events in the dark matter search region is about 200, giving a maximum in one PMT of $\sim100$ if the event is near the bottom or the top of the TPC. For the standard operating fields of the TPC, the S2 signal for electron recoils has typically $\sim30$ times more charge than the S1 signal but this charge is distributed over $\sim$ a few microseconds making the instantaneous rate from the S2 a less severe requirement than the S1 rate. The range beyond the dark matter region is also useful, however, for understanding the detector and estimating potential backgrounds. In practice, the system has a dynamic range of about 1,500 set by a combination of the cold pre-amplifier (see immediately below and section \ref{subsec:TPC-Preamplifier}) and the PMT. 

We experienced two challenges to achieving adequate signal to noise for single photo-electron signals. One was reducing the pickup on the 6 meter-long cables connecting the PMTs to the DAQ electronics, particularly from  instrumentation monitoring the cryogenics.  This challenge has been met with a careful system design, in particular design of the grounding and signal and power distribution system, and the choice of cables and connectors.  The second challenge was that, when operated in liquid argon at a gain above $10^{6}$ as required for single photo-electron identification,  the R11065 PMTs would occasionally start to produce tens of kHz up to a MHz of single photo-electron pulses.  We did not observe such behavior at room temperature and believe the effect may be due to some Malter-like effect \cite{ref:Malter} where charge builds up on a surface in the PMT which is conductive in the warm but insulating in the cold. The high-rate behavior is only cured by turning off the PMT HV for a period of several hours,  allowing the charge to be neutralized .  As described later, this second challenge has been met with the implementation of pre-amplifiers mounted directly on the PMT HV dividers in the liquid. These pre-amplifiers produce an effective gain of 24 compared to a conventional back-terminated 50 $\Omega$ system, and allow the PMTs to be run at a comfortable gain of $\sim 3 \times$ $10^{5}$.

A further important  requirement was that the TPC electronics in proximity to the argon not contribute any significant radioactive background. To this end, the elements of the electronics system in proximity to the TPC were part of a comprehensive radio-assay campaign designed to measure and minimize the radioactivity of the detector. From the beginning, it was recognized that radioactivity from the PMTs, particularly the stem region,  and from the cryostat vessel would dominate the radio-activity budget and the work on the electronics components was designed to ensure there were no unpleasant surprises. In terms of the most serious background, that from neutrons in the detector, the PMT electronics hardware is calculated to contribute less then 2\% of the total. Reference \cite{ref:WesterdaleThesis} has a complete table of the radio-assay results.



\section{Electronics System Overview}
\label{sec:Electronics-System-Overview}

\begin{figure}[h]
\begin{center}
\includegraphics[width=6.0in]{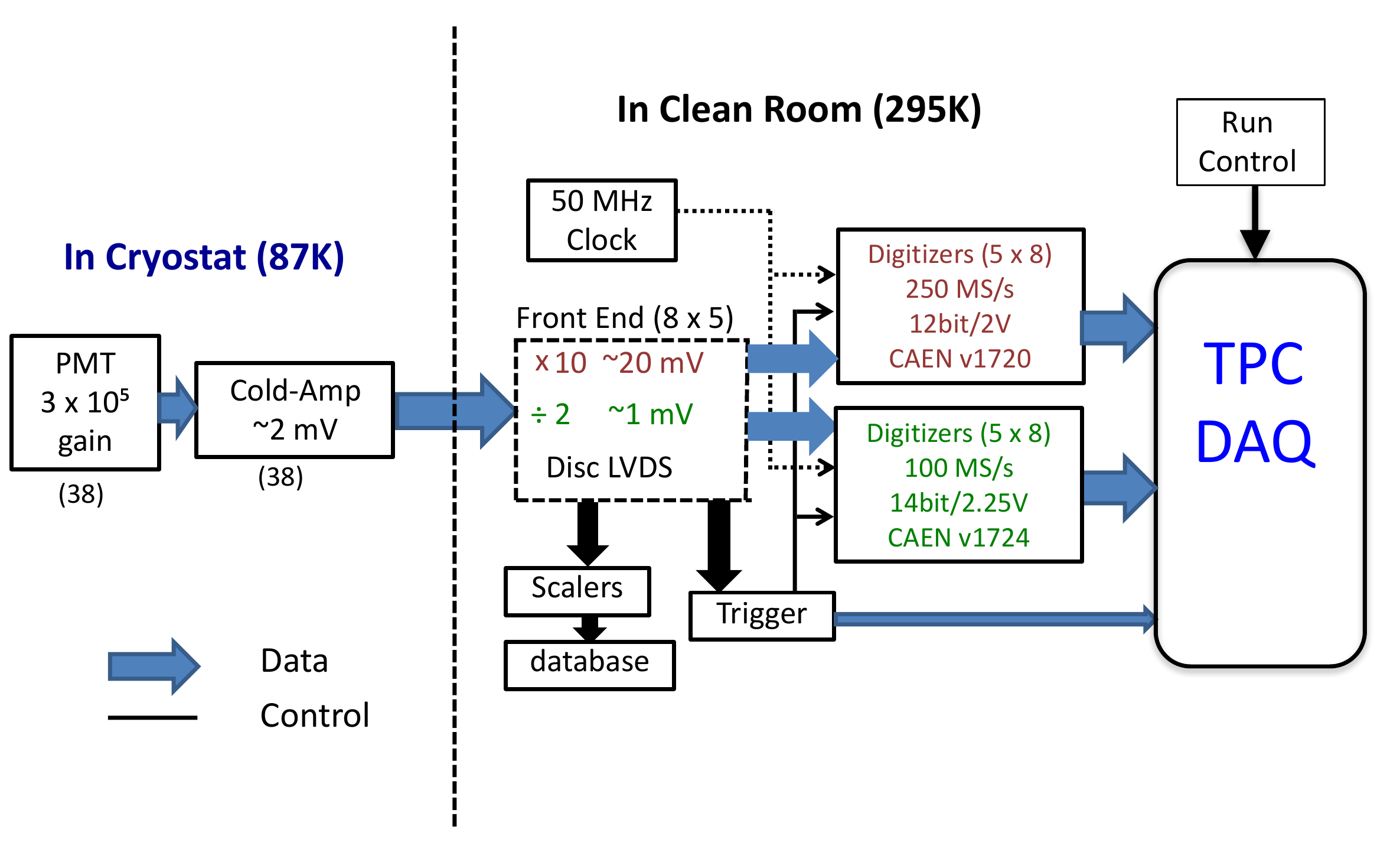}
\caption{A schematic of the TPC electronics chain}
\label{fig:TPC-electronics-schematic}
\end{center}
\end{figure}

Figure \ref{fig:TPC-electronics-schematic} shows the major items of the electronics chain. The current signal from each PMT passes directly into the pre-amplifier mounted on the HV divider. From the pre-amplifier, the signals pass along 7 m of cable to the Front End modules located in the clean room on top of the water tank. The Front End provides discriminated outputs for the trigger and the scaler recording system, and two analog copies of the input signal at two different gains for digitization in two sets of  waveform digitizers. The data are digitized and pushed to the data acquisition system on receipt of a trigger. A 50 MHz clock with its outputs arranged in `star' mode is used to synchronize the system. The stand-alone scaler system - independent of the main DAQ - records the rate of digitized pulses and is used for monitoring the health of the PMTs.

\begin{table}
\begin{center}
\small
\begin{tabular}{|llll |}
\hline\hline
 Function 	& Where &  Type & Requirement \\  
 \hline
Signal cable			& In cryostat	& Multiflex 86 (38)	& Double-shielded, small size (2.65mm) \\
Signal cable 			& External 	& H $\&$S K-02252-D  (38)		& Double-shielded \\
Preamp signal connector	& Preamp Board 			& MCX  (38)				& Quality,Size \\
Signal Feedthrough 		& WCD flange				& Accu-glass 100522 (38)	& Size, coax, vacuum tightness \\
Signal Input connector	& Front-end module			& SMA (38)				& Size, quality \\
HV cable				& In cryostat				& Accu-glass 100710 (38)	& Low radioactivity, HV   \\
HV cable				& External					& Connectronics AWG 24		& Rated for 2.5 kV   \\
HV feedthrough		& WCD Flange				& MDC Micro-Midgy	(40)	& Rated for 2.5 kV * \\
Preamp Power cable	& In cryostat				& Multiflex	 86 (2)		& Double-shielded	\\
Preamp Power cable	& External					& H $\&$S K-02252-D	& Double-shielded                   \\
Preamp Power feedthrough & WCD flange			& SMA (2)					&	\\
Preamp Power fanout  	&  Near PMTs				& Accu-glass 1000710	&  \\
\hline\hline
\end{tabular}
\caption{Cable and Connector Types used in TPC Signal and HV. \newline (* these required modification for use in argon gas)}
\label{tab:cables-and-connectors}
\end{center}
\end{table}
\normalsize

The signal and power lines of the TPC (at a temperature of 87K) pass from the TPC cryostat through the liquid scintillator veto tank and the water tank in a number of flexible stainless steel pipes to a set of feedthroughs on the top of the water tank in the clean room. To avoid the use of intermediate feedthroughs as the lines go between the liquid scintillator and the water tanks, the pipes have a rigid section welded to the top flange of the liquid scintillator tank.  

\begin{figure}[h]
\includegraphics[width=0.30\textwidth]{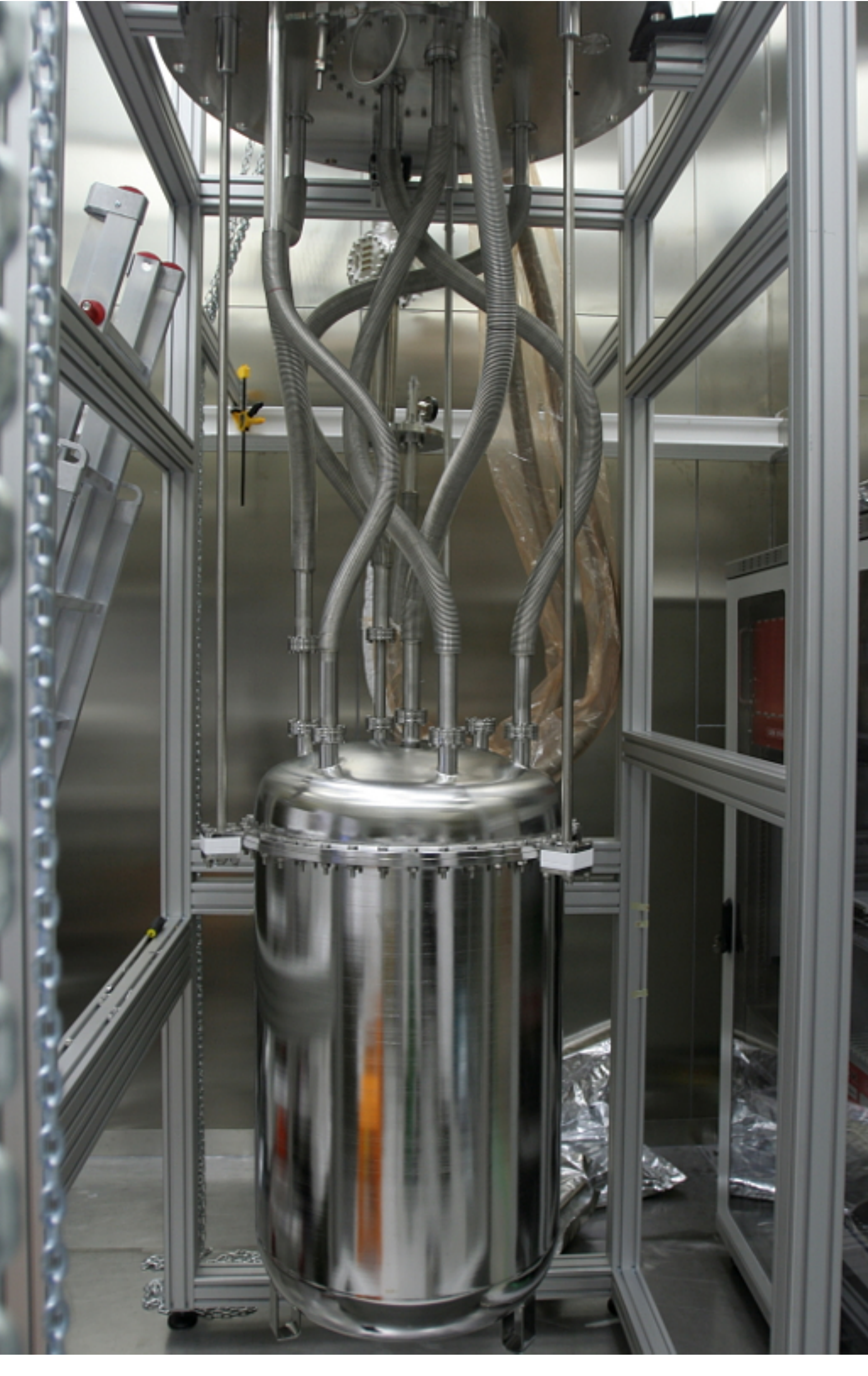}
\includegraphics[width=0.67\textwidth]{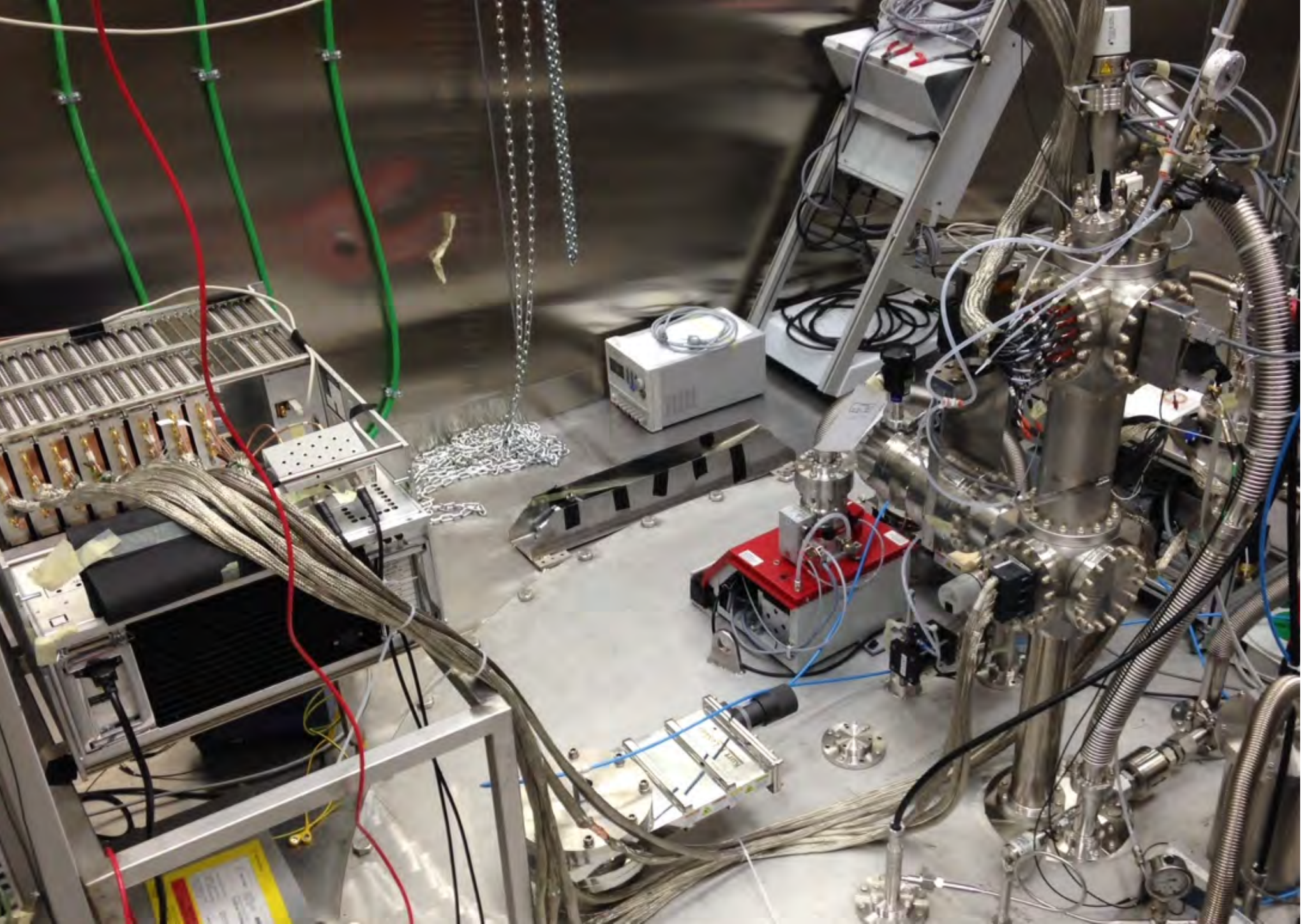}
\caption{Left: The TPC cryostat in the cleanroom with its piping and the flange of the LSV.\\
  Right: At the left, the front-end modules and inputs and at the right the feedthrough column (the `squid head')  for the signal cables and PMT high voltage cables.}
\label{fig:TPC-feedthroughs}
\end{figure}

Figure \ref{fig:TPC-feedthroughs} shows the flexible pipes that connect the TPC cryostat through the top-flange of the LSV, and the feedthroughs where the signal and PMT HV and pre-amp power cables emerge into the electronics room.

The primary concern throughout the system design and implementation was to reduce noise, both internally generated and from pick-up from external sources, to a minimum. This concern drove the choice of connectors, cables and the `grounding' scheme. Early tests showed the advantages of double-shielded cables in reducing pick-up and as a result the signal cables are double-shielded along their entire path from the TPC to the Front-End-Modules in the Clean room.  The cables and connectors are given in Table \ref{tab:cables-and-connectors}.
To avoid ground loops, a single current-return point (`ground') was established for the PMT signals by the use of insulated-ground coaxial feedthroughs. The sealed metal cryostat provides shielding from external noise sources but penetrations for test and instrumentation cables could and did convey noise into the cryostat.  Pump-drivers and switching power-supplies were the major culprits. 

Pump motor-controllers were especially troublesome sources because most sent unfiltered and un-shielded signals to the appliance they controlled. These sources of noise were mitigated by using variac-based sinusoidal motor controls. Noisy switching power supplies were identified and treated by adding filtering or replacing the supplies with linear components. Noise from the PMT high voltage (HV) supplies was reduced by placing filters on each HV output.  Ground loops were avoided by using isolated feedthroughs. At the same time, all cable shields were connected via low-inductance AC coupling to the cryostat surface to reduce the RF energy which could otherwise enter the cryostat along the shields. As a result, the noise at the output of the amplifier chain is dominated by the pre-amplifier electronic noise floor and there are no significant spectral components from external noise within the frequency range of interest. The output noise level achieved is 70 $\mu$V (rms) over a bandwidth of 150 MHz.

\begin{center}
\begin{figure}[h]
\includegraphics[width=0.95\textwidth]{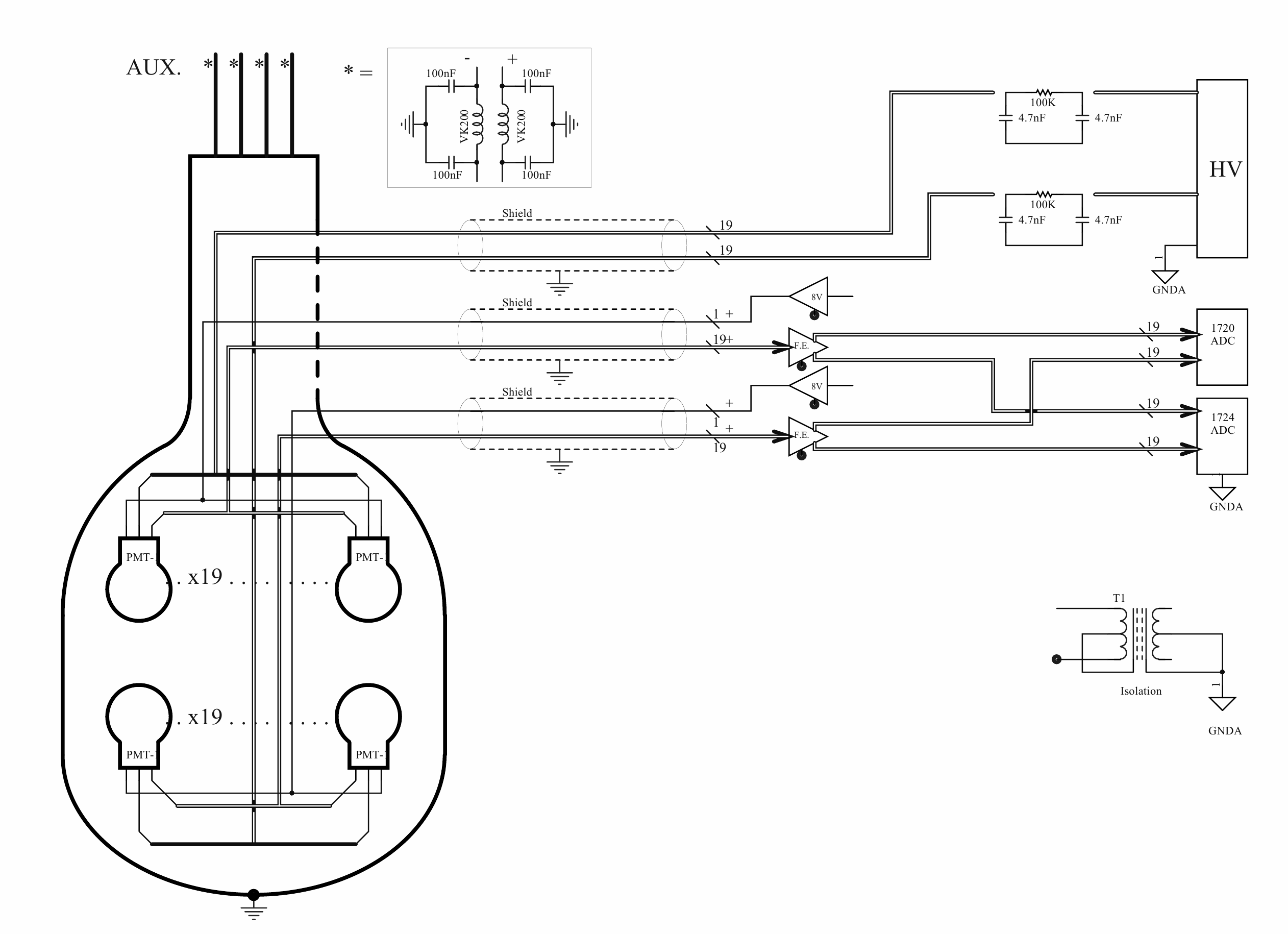}
\caption{Line drawing showing the grounding arrangements. The TPC, its support structure including the clean room structure and floor, are considered the primary `ground', shown by the standard ground symbol. The utility power is shown as GNDA. The pre-amplifier power supplies and the front-end modules were on a separate distinct ground of their own.}
\label{fig:TPCLine-drawing}
\end{figure}
\end{center}

Figure \ref{fig:TPCLine-drawing} is a sketch of the signal and power distribution scheme. The PMT HV returns are connected to ground at the PMT HV supply and with 500 $\Omega$ resistors to the return of the PMT signal at the PMT. As they leave the PMT HV supply, the PMT HV cables are routed in two bundles of 19 cables, each bundle enclosed inside a copper braid till it reaches the feedthrough. At the feedthrough where the HV cables pass from the clean-room environment to the environment of the metal tubes connected to the cryostat, (see Figure~\ref{fig:TPC-feedthroughs} right) the returns of the 19 cables in a single bundle are tied together and pass through a single isolated connection. A 40 channel feedthrough is used to take the 38 HV lines and the two returns. The PMT HV supply itself provides the HV and a bothersome switching noise with a period of about 4 $\mu$s. To remove this noise, each HV line is filtered with a C-R-C filter where C = 4.7 nF and R = 10k$\Omega$. 

Two cables are used to provide power from the clean room to the pre-amplifiers on the PMT HV dividers; each cable serves a set of 19 channels (one set at the top and one set at the bottom of the TPC). The 38 signal cables from the pre-amplifiers and the two pre-amplifier power lines are bundled in two groups, each group made up of four sets of five cables (19 signal and 1 power line) and each set, again, is inserted into copper braid. These lines pass from the clean room to cryostat environment through accu-glass coaxial feedthroughs.  The arrangement with a single cable for the power and return to the pre-amplifiers of 19 PMT channels and the pre-amplifier signals returns tied together at the Front-End-Module introduces a ground-loop, but one of small area.   The return of the signals and the preamp power are isolated from the general ground by an isolation transformer which powers the signal receivers (Front-End-Modules) and the pre-amplifier power supply. 

The line drawing also shows a C-L-C filter schematic. Such filters are used on the instrumentation signaling, labeled AUX in the drawing, where a series resistance and therefore a C-R-C filter could not be implemented. In this case C = 100 nF, and the L is provided by a Ferroxcube VK200. 


\section{TPC electronics custom components}
\label{sec:TPC-Electronics-Custom-Components}
The custom components of the TPC electronics are :
\begin{itemize}
\itemsep0em
\item the PMT base
\item the pre-amplifier 
\item the front-end amplifier and discriminator,  and the scaler recording system
\item the clock fanout used to synchronize the waveform digitizers
\end{itemize}

\subsection{PMT Base}
\label{subsec:PMT-Base}

The PMTs are equipped with a negative HV resistive divider using surface-mount metal-film resistors and COG capacitors; both types of components have small temperature coefficients (< 3\% between ambient and 87K). The circuit board is constructed in Cirlex~\cite{ref:Cirlex} because of its low radioactivity and metal-like coefficient of thermal expansion.  Low activity solder in paste form~\cite{ref:Solder} was used in the assembly. 
 
\begin{figure}[h]
\includegraphics[width=6.0in]{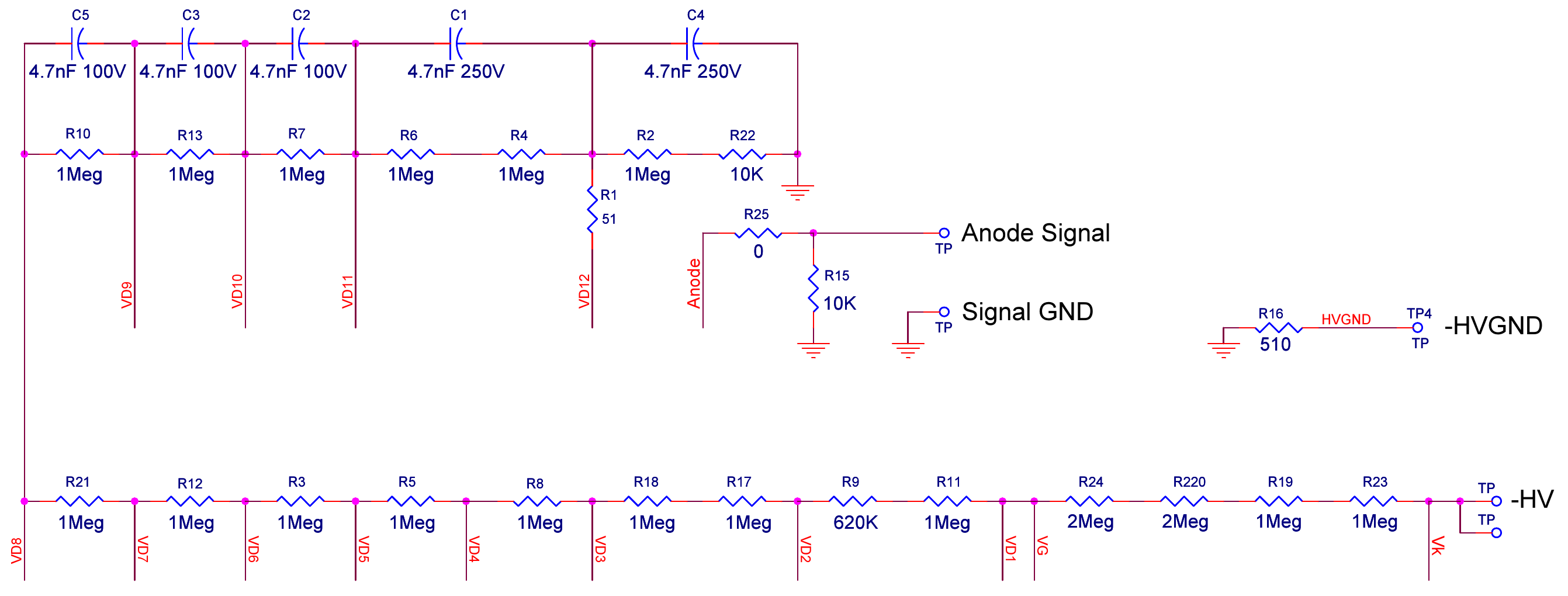}
\caption{The PMT base schematic}
\label{fig:PMT-Base-Schematic}
\end{figure}

Because we are running the PMTs at low gain and therefore lower (absolute) voltages ($\sim$-1200 Volts) than typical for these tubes, tests were made to establish the minimum voltage between the photo-cathode and first dynode required for maximum collection efficiency; these tests showed that the maximum collection efficiency is achieved once the potential difference is at least 300 Volts.  As a consequence, the divider ratios recommended by Hamamatsu were modified to put a larger fraction (29\% cf 21.5\%) of the total voltage between the photo-cathode and the first dynode, giving a typical potential difference between photo-cathode and first dynode of 350 Volts. Figure \ref{fig:PMT-Base-Schematic} shows the divider schematic. 
 
The advantage of running with negative HV is that it allows for DC coupling to the anode signal and avoids baseline shifts inherent in AC coupling that affect integration of long signals. One disadvantage of using -HV with these particular PMTs is that the enclosing bulb is metal and is maintained at cathode potential, thereby putting a large metal surface at -HV.  Another observation was that some of the tubes were sensitive at turn-on to the potential of other surfaces near the photo-cathodes, taking of order an hour to reach full gain.  The outer surfaces of the silica windows were coated with ITO as an electrostatic shield for the PMTs and this turn-on effect was mitigated by maintaining the ITO at the average potential of the PMT cathodes facing them.
\subsection{TPC Preamplifier}
\label{subsec:TPC-Preamplifier}

Early experience of the R11065 photomultipliers at cryogenic temperatures showed that, after some hours or a few days of operation in the cold and at the gain required for clean single photo-electron detection, the PMTs would go into a mode of continuous discharge - with up to a MHz single photo-electron rates, and the emission of light. The PMT would only recover after its HV was turned off for several hours. To mitigate this behavior which was observed to depend on the PMT gain, a cryogenic pre-amplifier has been developed. The preamplifiers mount directly on the PMT bases in the liquid argon, allowing the PMTs to run at reduced gain.  Their implementation has reduced the occurrence of this behavior to less than one incident per month.
 
\begin{figure}[h]
\begin{center}
\includegraphics[width = 8.cm ]{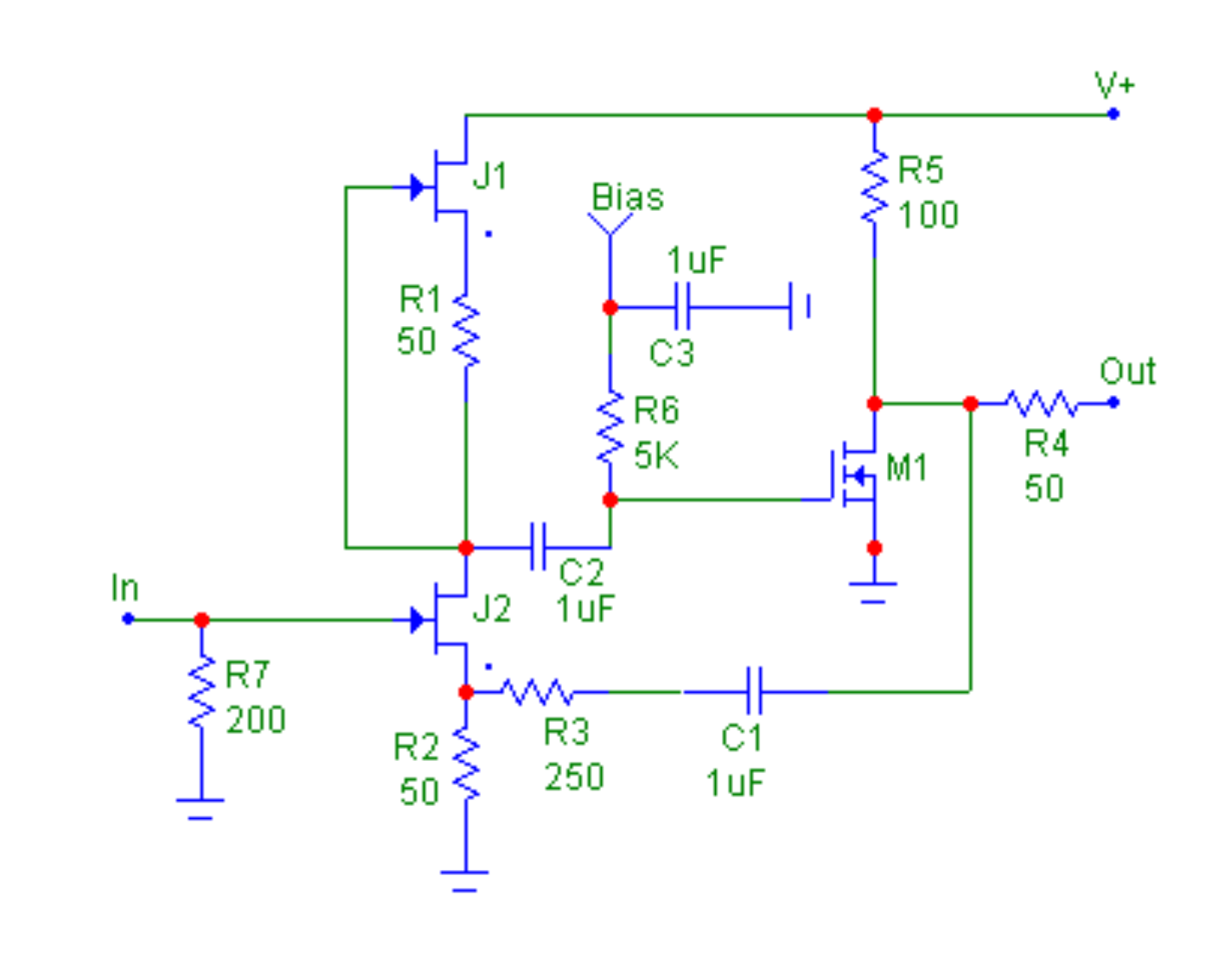} 
\includegraphics[width = 7.cm ]{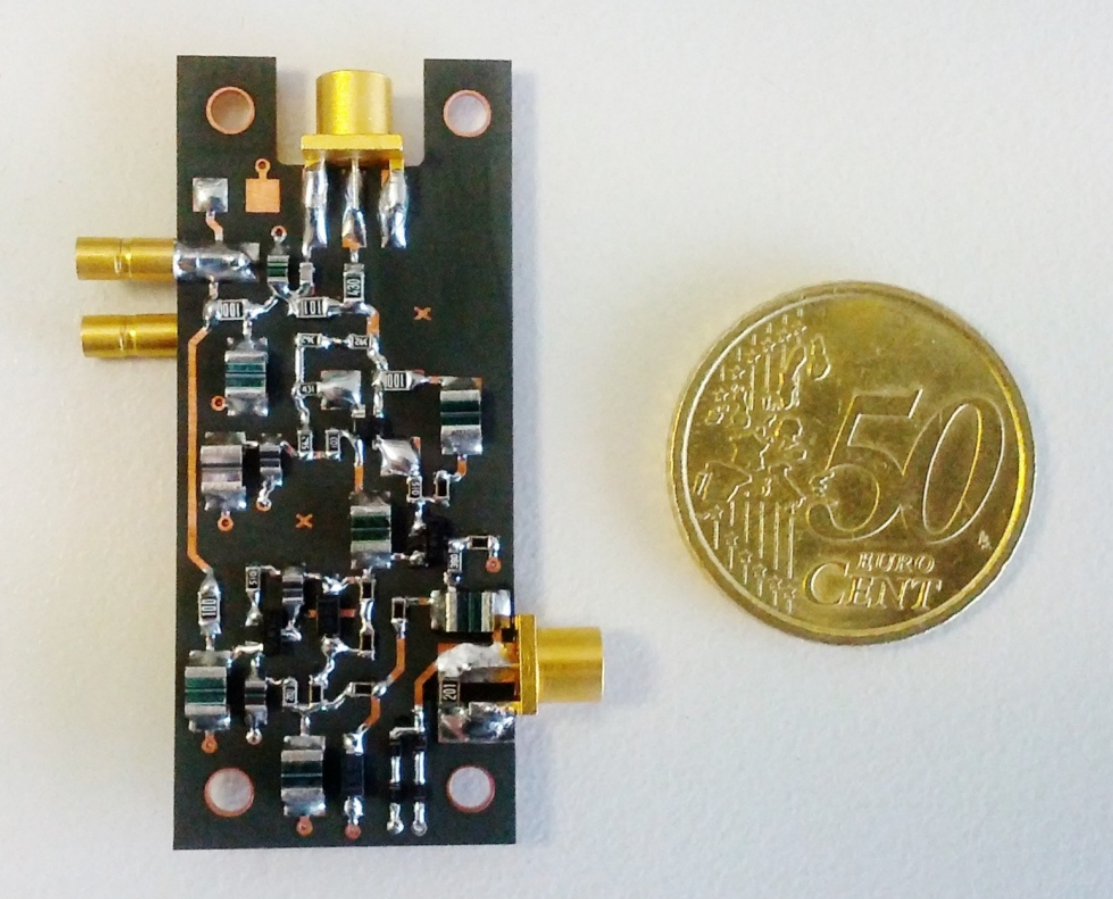}
\caption{The cryogenic pre-amplifier; Left: A simplified schematic with indicative values of the components; Right: A photograph of the cryogenic pre-amplifier.}
\label{fig:TPCPreamp-circuit-and-photo}
\end{center}
\end{figure}

Details of the design considerations and choices and the complete design, including all part numbers, are to be given elsewhere; we give a summary of the main considerations in the design and of the performance.  Figure ~\ref{fig:TPCPreamp-circuit-and-photo} shows a schematic of the pre-amplifier and a photograph of the actual device.  

The circuit is made of discrete components chosen for specific aspects of their performance; low-noise was an important criterion for the input JFETs and output drive transistor, low radioactivity and small temperature change of the capacitance from room to cryogenic temperatures were criteria for the capacitors. The circuit is constructed on a Cirlex ~\cite{ref:Cirlex} substrate, the same material as used for the PMT HV dividers, in order to avoid mechanical stress between the pre-amplifier and the divider on which the preamplifier is mounted.  A feature of the circuit design is that its operation is quite similar from room temperature to liquid nitrogen - thereby allowing developments to be made conveniently at room-temperature.    

Limiting the power consumption to avoid generating bubbles in the argon was another key consideration in the design. The pre-amplifier operates with a power consumption of less than 90 mW/channel (12 mA at 7V). 
The circuit takes advantage of the single-sided (negative) nature of the PMT signals to place the quiescent level of the output near the positive rail, thereby reducing the power consumption. This leads to putting a  capacitor (C1) in the feedback loop to decouple the DC offset of the output, thereby introducing a lower cut-off frequency in the gain. Rather than compensate for this feature in the pre-amplifier circuit itself  - compensation which would require large, radio-pure and cryogenically capable capacitors - the compensation is performed at the input to the front-end module in the clean-room where the physical constraints of space, temperature dependence, and radio-purity are much relaxed.  

\begin{figure}[h]
\begin{center}
\includegraphics[width = 7.cm]{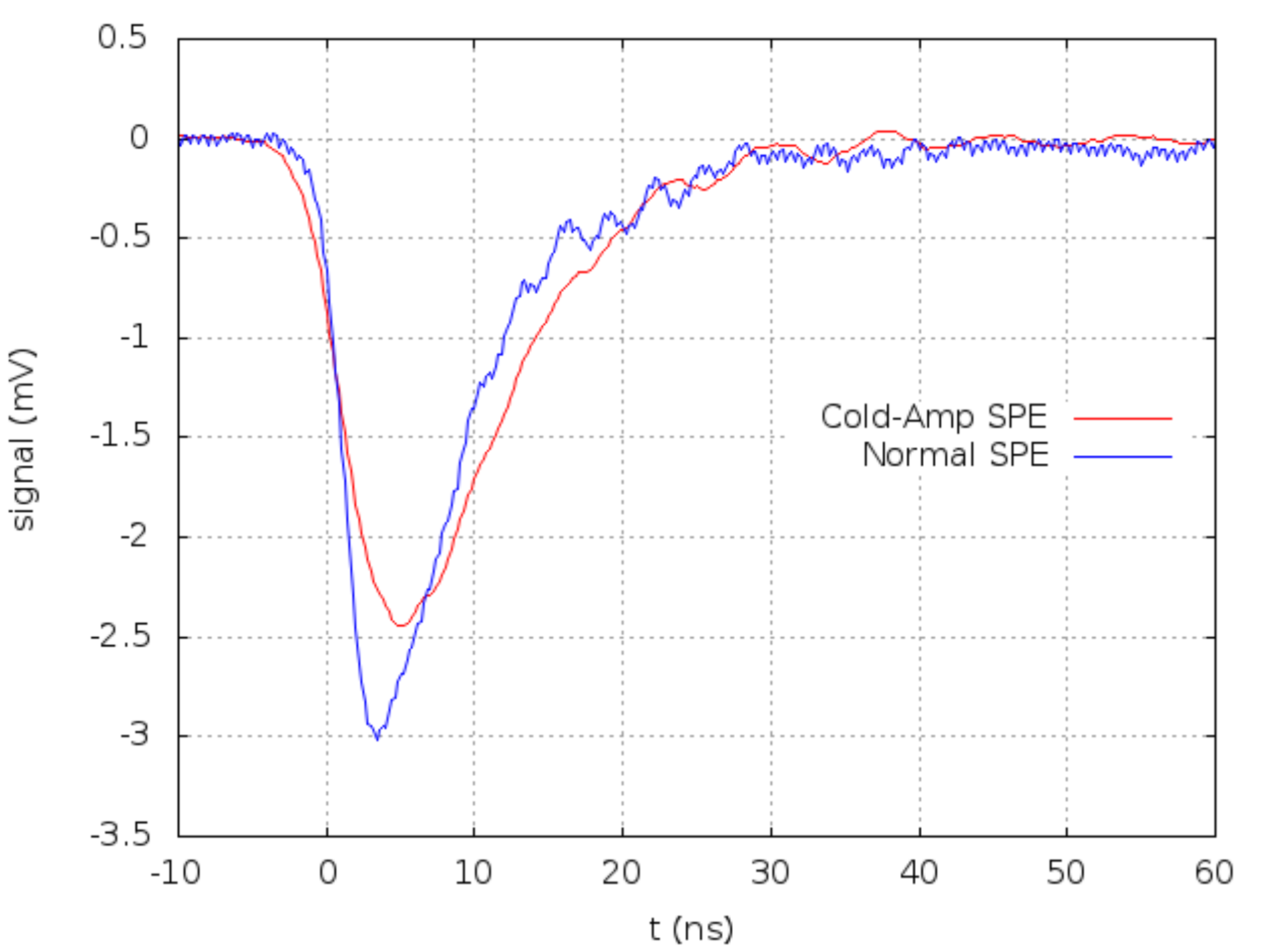}
\caption{The pre-amplifier output pulse for a single photo-electron compared with a single photo-electron directly from the PMT}
\label{fig:TPCPreamp-spe}
\end{center}
\end{figure}

The PMT is a current source and the pre-amplifier is mounted directly on the PMT base;  there is, therefore, some freedom in determining the pre-amplifier input impedance and there is an advantage for signal voltage in using a higher impedance, if possible, than is used in a conventional 50 $\Omega$ system. A value of 200 $\Omega$ for the input impedance was determined by an optimization of signal-to-noise and waveform fidelity. (Higher values of the input resistance give better signal-noise but reduce the signal fidelity.) The gain of the pre-amplifier circuit itself is 6 and it operates into a back-terminated 50  $\Omega$ system as shown in Figure ~\ref{fig:TPCPreamp-circuit-and-photo}; half the output voltage appears across the back termination (R4 in the figure), and the other half across the 50 $\Omega$ of the receiving system, the front-end module. The signal voltage seen at the front-end module for a current $\mathcal{I}$ out of the PMT is, therefore, 200 $\times$ 6/2 $\times$ $\mathcal{I}$ = 600 $\times$ $\mathcal{I}$. We can compare this to a back-terminated 50 $\Omega$ situation without the pre-amplifier where the voltage across the receiving module would be 25 $\times$ $\mathcal{I}$; the effective  gain due to the pre-amplifier is therefore  600/25 = 24.  Figure \ref{fig:TPCPreamp-spe} shows the output of the pre-amplifier compared with the output directly from the PMT base for a single photo-electron.The capacitance of the PMT anode in series with the 200 $\Omega$ input of the pre-amplifier acts as a shaper and results in a completely acceptable slight rounding of the pulse. The maximum output is 3V and with an output of typically 2 mV from a single photo-electron , the system provides a dynamic range of 1500 photo-electrons.

\begin{figure}[h]
\begin{center}
\includegraphics[width= 2.5 in ]{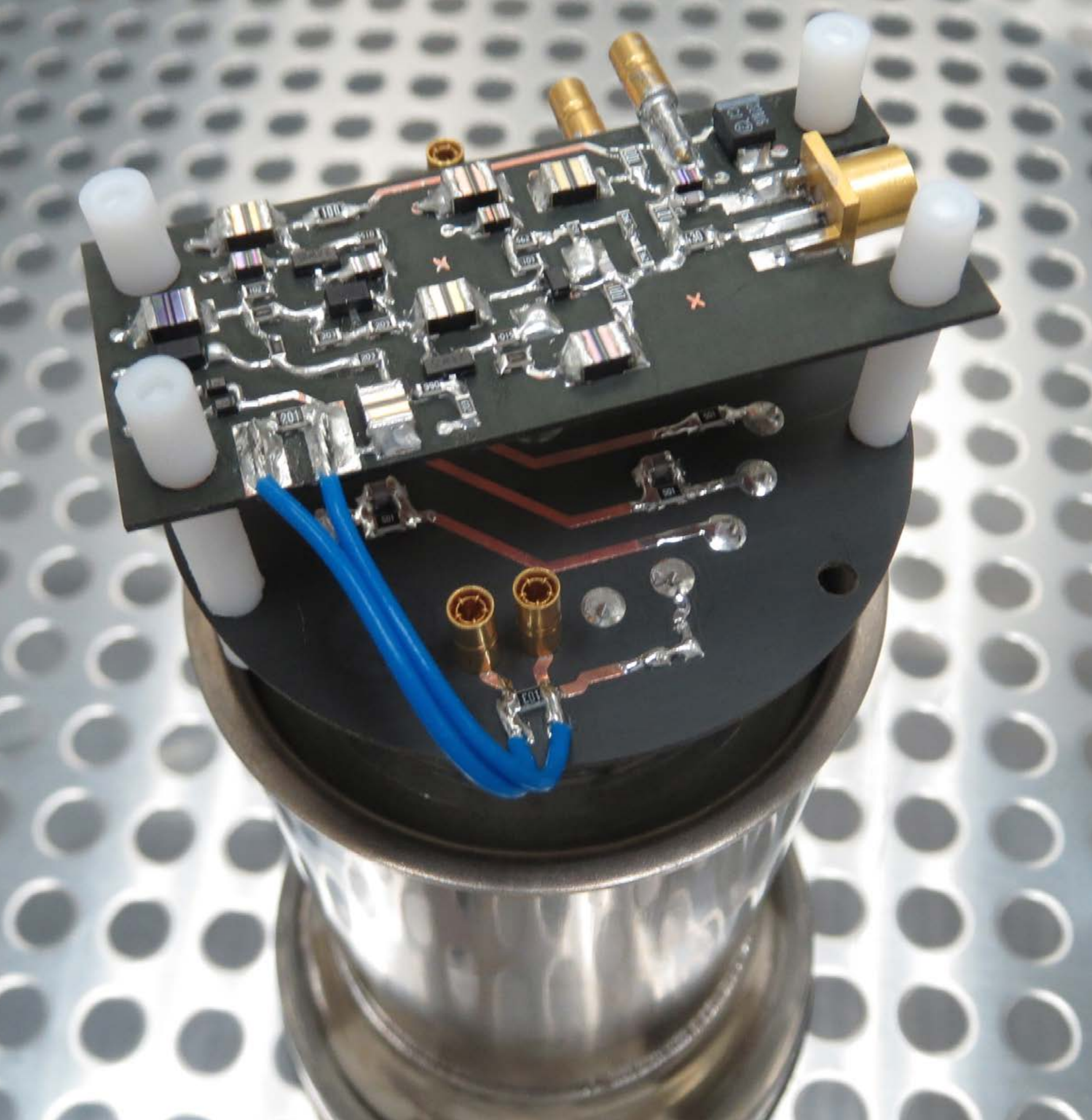}
\includegraphics[width= 3. in ]{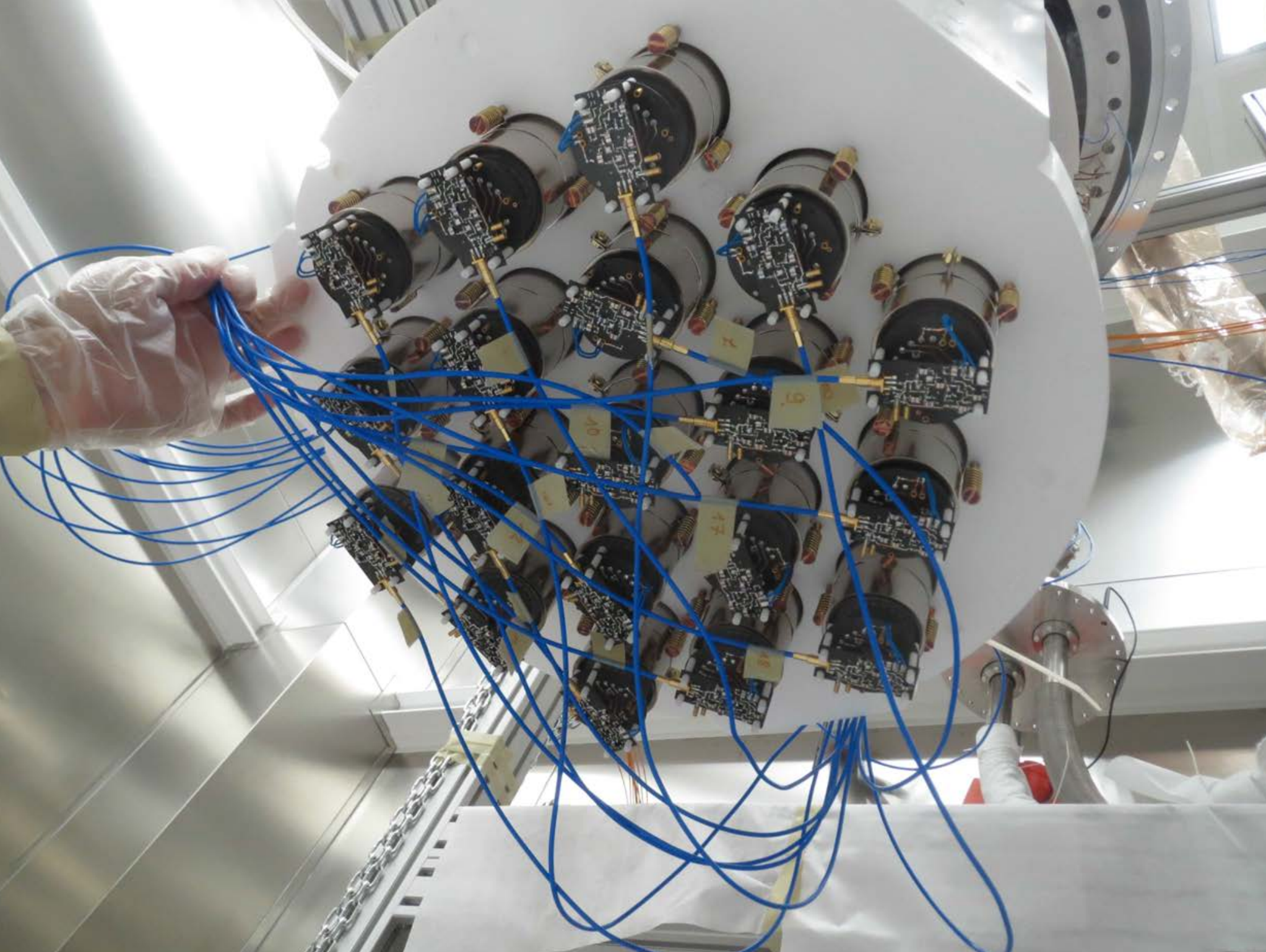}

\caption{left: the cold pre-amplifier mounting; right: the lower array of PMTs with the pre-amplifiers and signal cables}
\label{fig:lower-array}
\end{center}
\end{figure}

Figure ~\ref{fig:lower-array} shows the mounting of the pre-amplifier on the PMT base and the lower array of the PMTs with bases and pre-amplifiers with the signal cables connected.


\subsection{TPC Front-End}
\label{subsec:TPC-Front-End}

The custom ``Front-End''  modules are located in the clean room above the experiment water tank as close to the point where the LArTPC signals emerge as practical. They receive the signals from the cold pre-amplifiers and provide the logic signals used to form the trigger,  the analog outputs that go to the high and low gain waveform digitizers, an analog output for debugging, and the logic signals used to record the PMT rates. Specifically, each channel provides:
\begin{itemize}
\itemsep0em
\item a $\times 10$  signal with 100 MHz shaping time 
\item a $\times 10$ full bandwidth signal 
\item a $\times 0.5$ signal with 40 MHz shaping time
\item two time-over-threshold LVDS discriminated outputs with individual channel thresholds settable both externally and internally
\end{itemize}

\begin{figure}[h]
\includegraphics[width=0.55\textwidth]{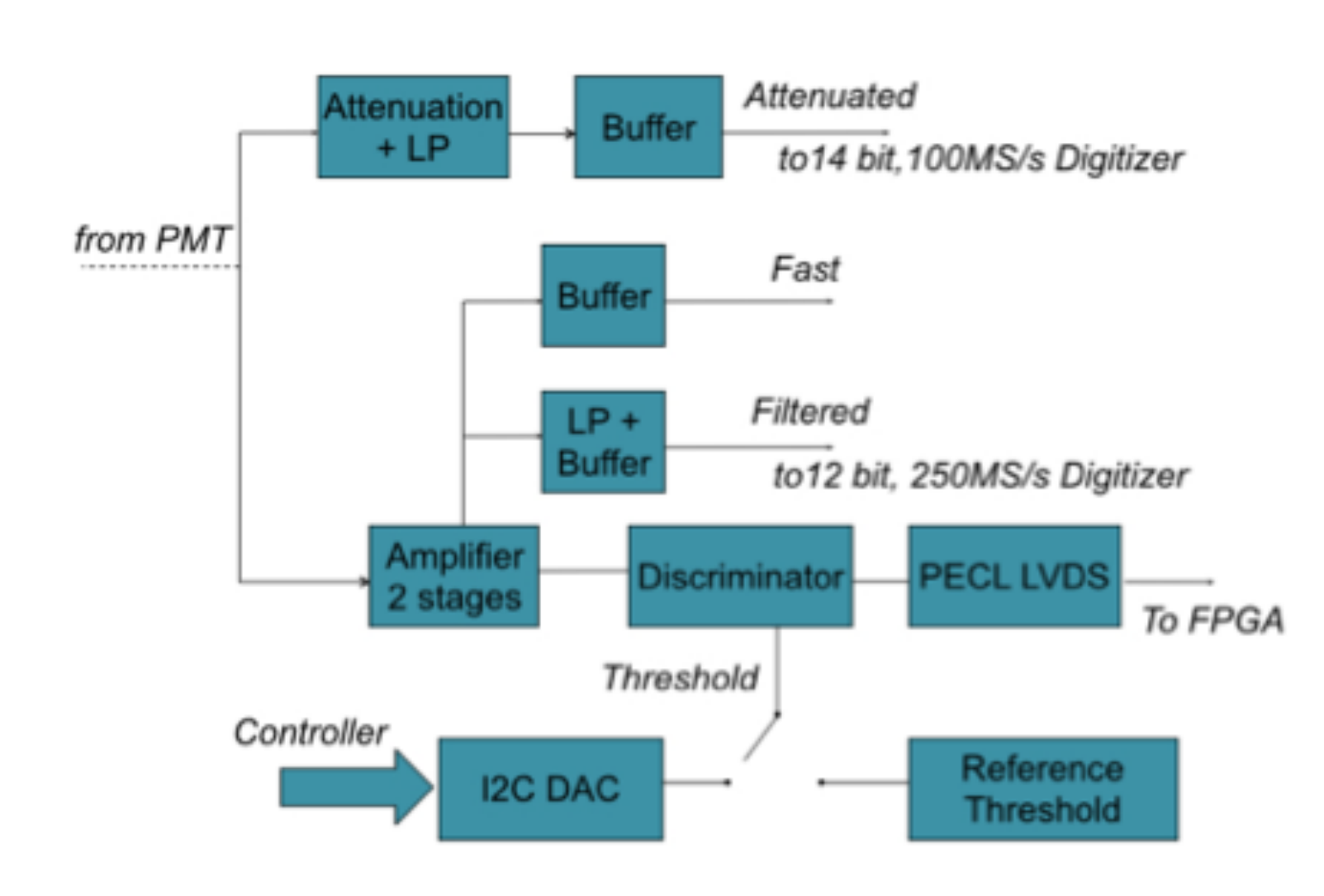}
\includegraphics[width=0.40\textwidth]{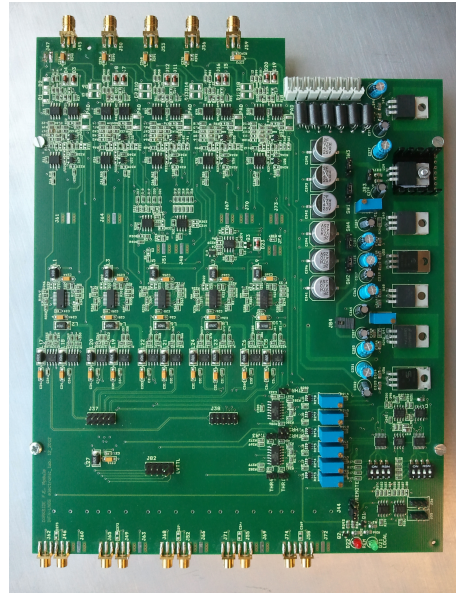}
\caption{left: A block diagram of the Front-End-Module; right A photograph of the module}
\label{fig:FEMBlock-Diagram}
\end{figure}

The module is realized in single-width NIM with 5 channels/unit. Details of the design and the extensive measurements made on the performance of the various outputs  are available in \cite{ref:Front-End-Thesis}. Figure \ref{fig:FEMBlock-Diagram} shows a block diagram and a photograph of the module board, and Table \ref{tab:Front-End-Performance} gives the values of some of the  performance parameters. 

The analog  output signals are buffered to isolate them from the inputs. The full bandwidth output is used for oscilloscope inspection.  The shaping times on the other analog outputs are set to match the bandwidth of the receiving digitizers; the receivers for the $\times 10$ and the $\times 0.5$ signals sample at 250 MS/s and 100 MS/s  respectively. The threshold for the LVDS (discriminated) outputs is typically set at 60\% of the single photo-electron peak amplitude or 1.2 mV at the input. One set of the LVDS signals is used to form the experiment trigger and the other is used to measure the rates of all PMTs as an online monitor. The minimum LVDS output width of 4 ns matches the latching requirements of the CAEN V1495 used to form the trigger. The choice of components was determined by a careful consideration of the bandwidth and noise requirements with an emphasis, again, on keeping the noise to a minimum.

\begin{figure}[h]
\includegraphics[width=0.48\textwidth]{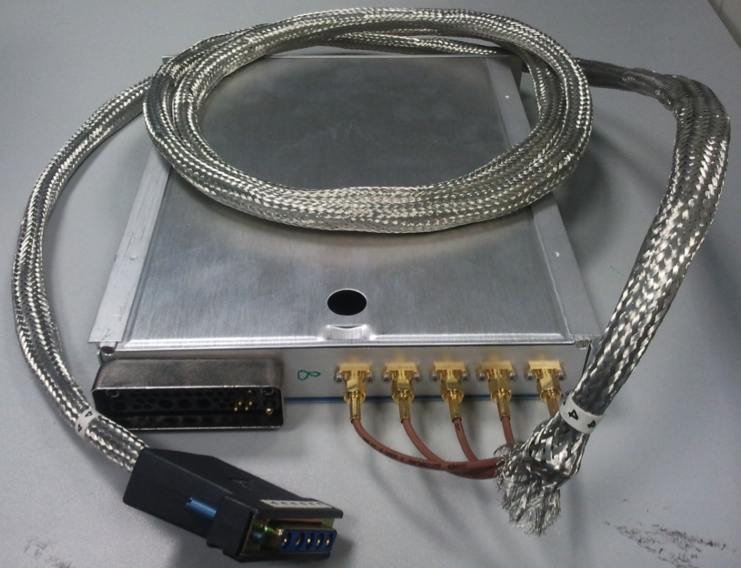}
\includegraphics[width=0.48\textwidth]{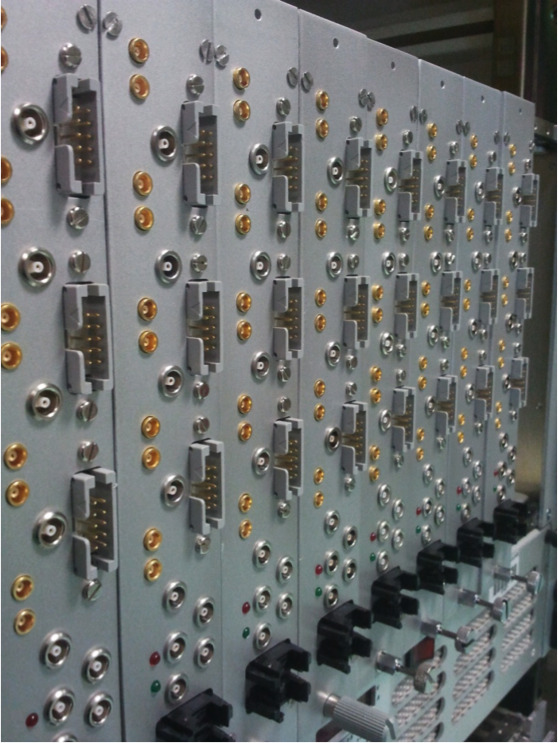}
\caption{Left: A complete module with the inputs and connections to the feedthroughs; Right; the front panels with the analog and LVDS outputs }
\label{fig:FEMCompletes}
\end{figure}

Figure \ref{fig:FEMCompletes} shows the complete module with its five inputs matching the signal feedthroughs and the front-panel connectors for the various outputs.

\begin{table}
\begin{center}
\small
\begin{tabular}{|ccccc |}
\hline
 Output	& Gain & Bandwidth (MHz) & Rise Time (ns) & RMS Noise (${\mu}V)$ \\  
 \hline
Full Bandwidth			& 10		& 282 	& 1.61               & 247 \\
100 MHz Shaping		& 10	 	& 97		& 3.37 	   	 & 196 \\
40 MHz Shapiing		& 0.5	& 50		& 7.29		 & 82 \\
\hline\end{tabular}
\caption{Bandwidth and noise performance of the Front-End module}
\label{tab:Front-End-Performance}
\end{center}
\end{table}

\normalsize

\begin{figure}[h]
\includegraphics[width=0.9\textwidth]{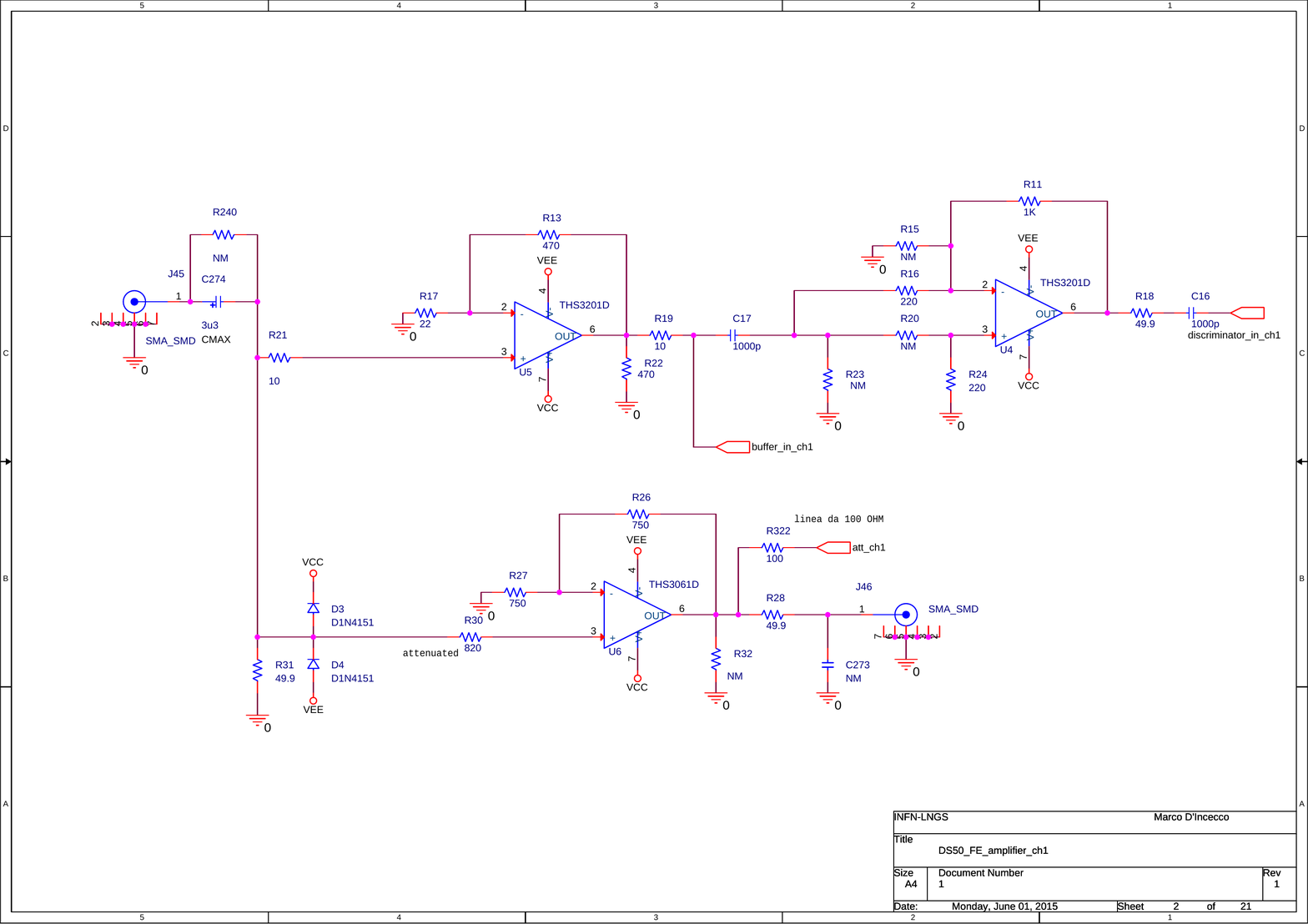}
\caption{Schematic of the Front-end-module amplifier section}
\label{fig:TPCFEM-Amplifier}
\end{figure}

\begin{figure}[h]
\includegraphics[width=0.9\textwidth]{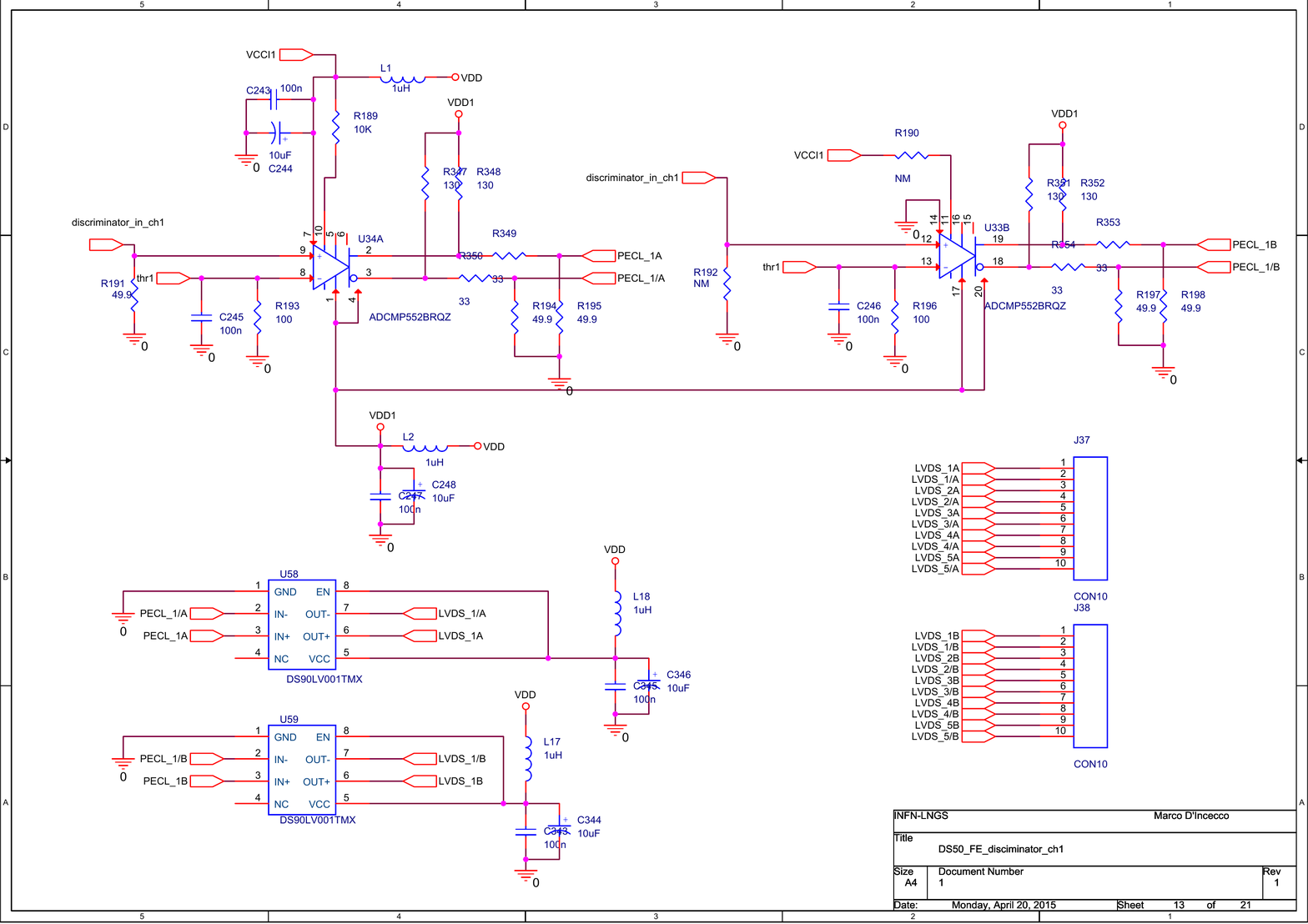}
\caption{Schematic of the Front-End-Module discriminator section}
\label{fig:TPCFEM-Discriminator}
\end{figure}

Figures \ref{fig:TPCFEM-Amplifier} and \ref{fig:TPCFEM-Discriminator} give the schematics of the fast amplifier and discriminator sections of the Front-end module. 

\begin{figure}[h]
\includegraphics[width=0.9\textwidth]{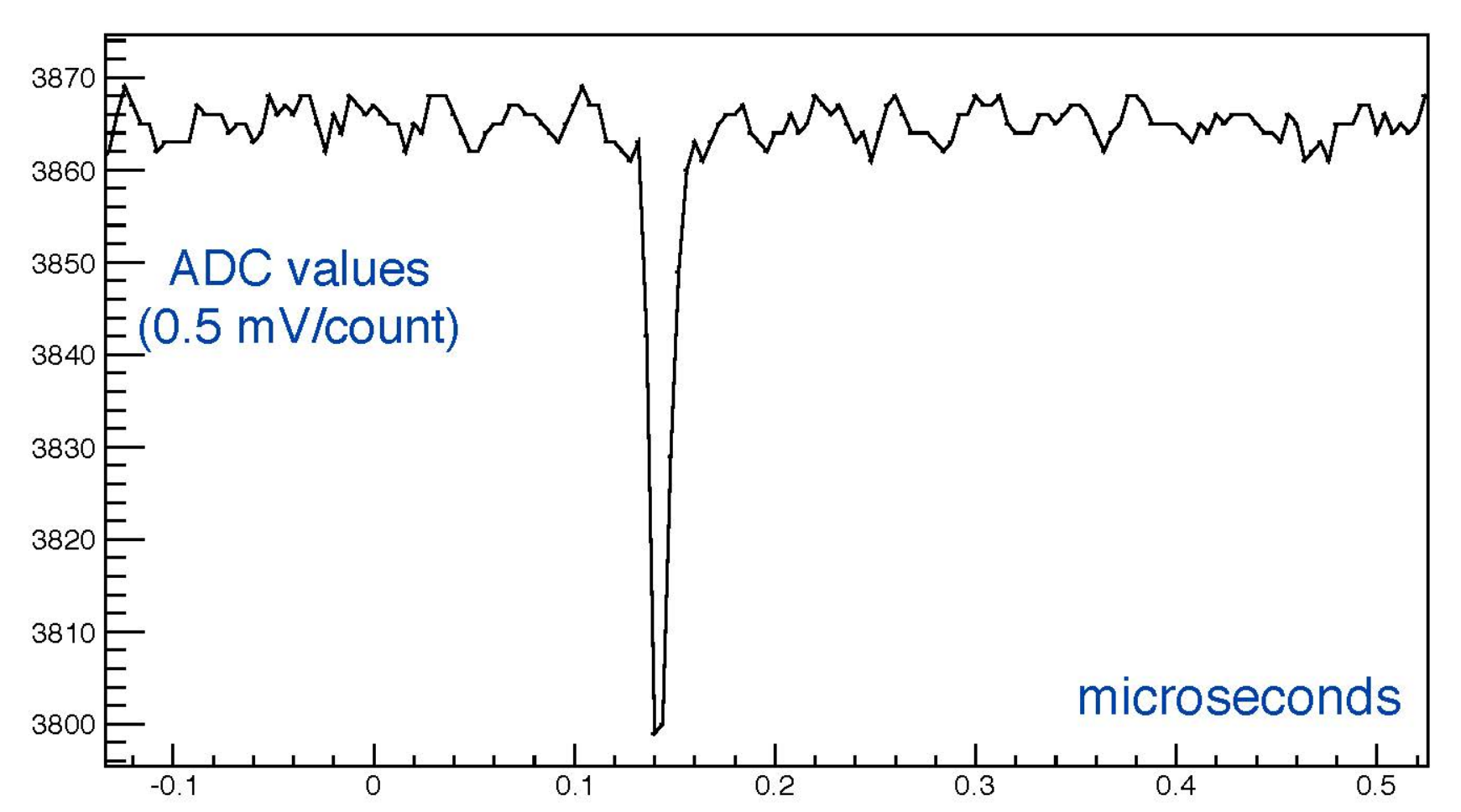}
\caption{A single photo-electron as digitized in the CAEN V1720 waveform digitizer. }
\label{fig:TPCSPE-in-digitizer}
\end{figure}

Figure \ref{fig:TPCSPE-in-digitizer} shows a single photo-electron output as digitized in the CAEN V1720 modules. The typical S/N is better than 30:1.

\subsection{TPC Front-End Scalers}
\label{subsec:TPC Front-End Scalers}
A convenient addition to the the Front-End electronics is a system of 40 scalers which records the rates of discriminator outputs from each of the individual channels and allows for the monitoring of the PMT rates independent of the main data-acquisition. The system is based on a commercial FPGA module \cite{ref:Opalkelly}  and uses one of the two discriminator outputs available for each channel as input. The FPGA counts the pulses for each channel for a fixed window, set originally to  0.3 s and then extended to 1 s for the low-radioactivity argon data, with a refresh period of 5 s. These rates are directly available to the shifter via the slow control display. The system is linked to the experiment database and also allows long-term monitoring of PMT behavior.


\subsection{TPC Clock Module}
\label{subsec:TPC-Clock-Module}

\begin{figure}[h]
\includegraphics[width=3.0in]{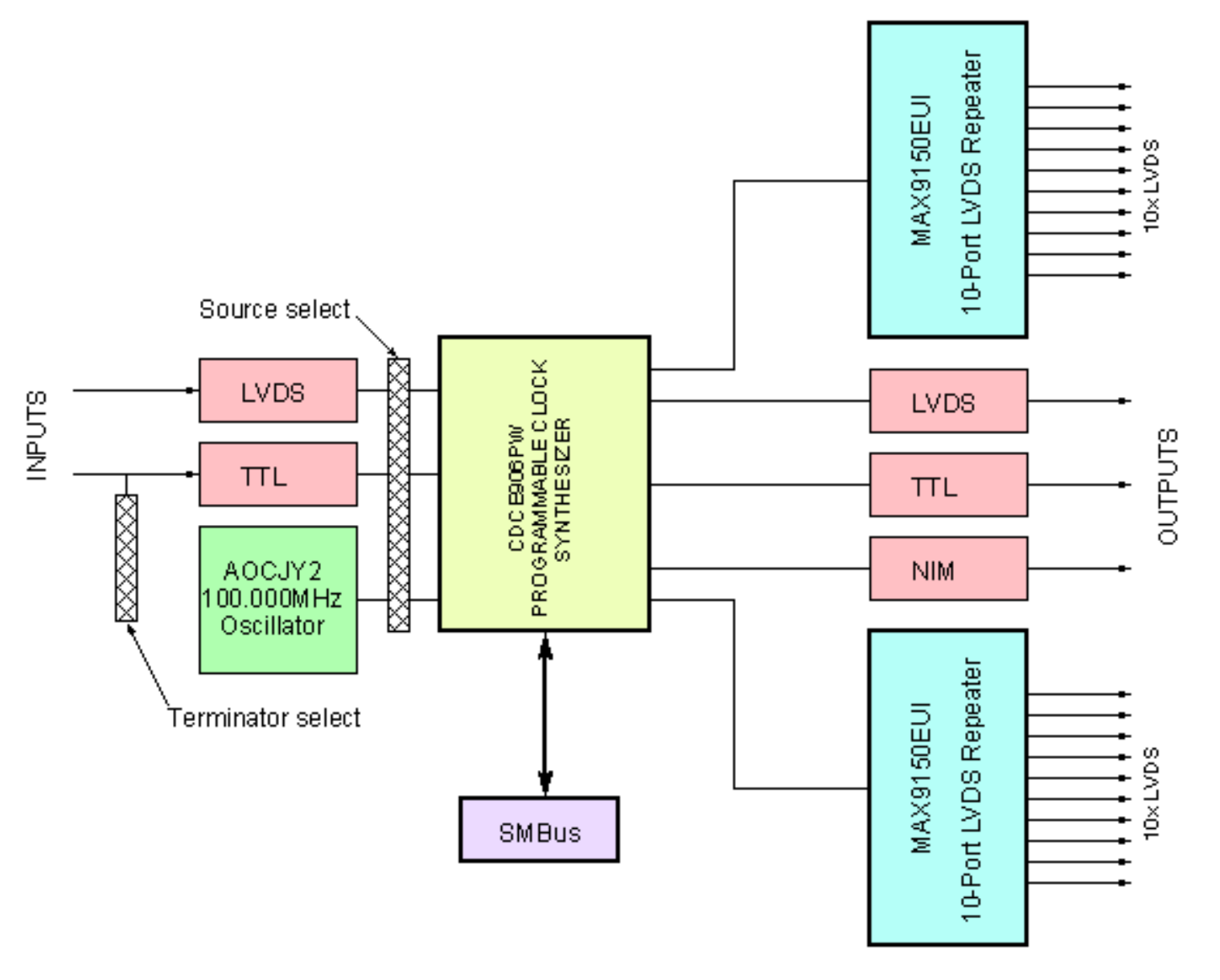}
\includegraphics[width=3.0in]{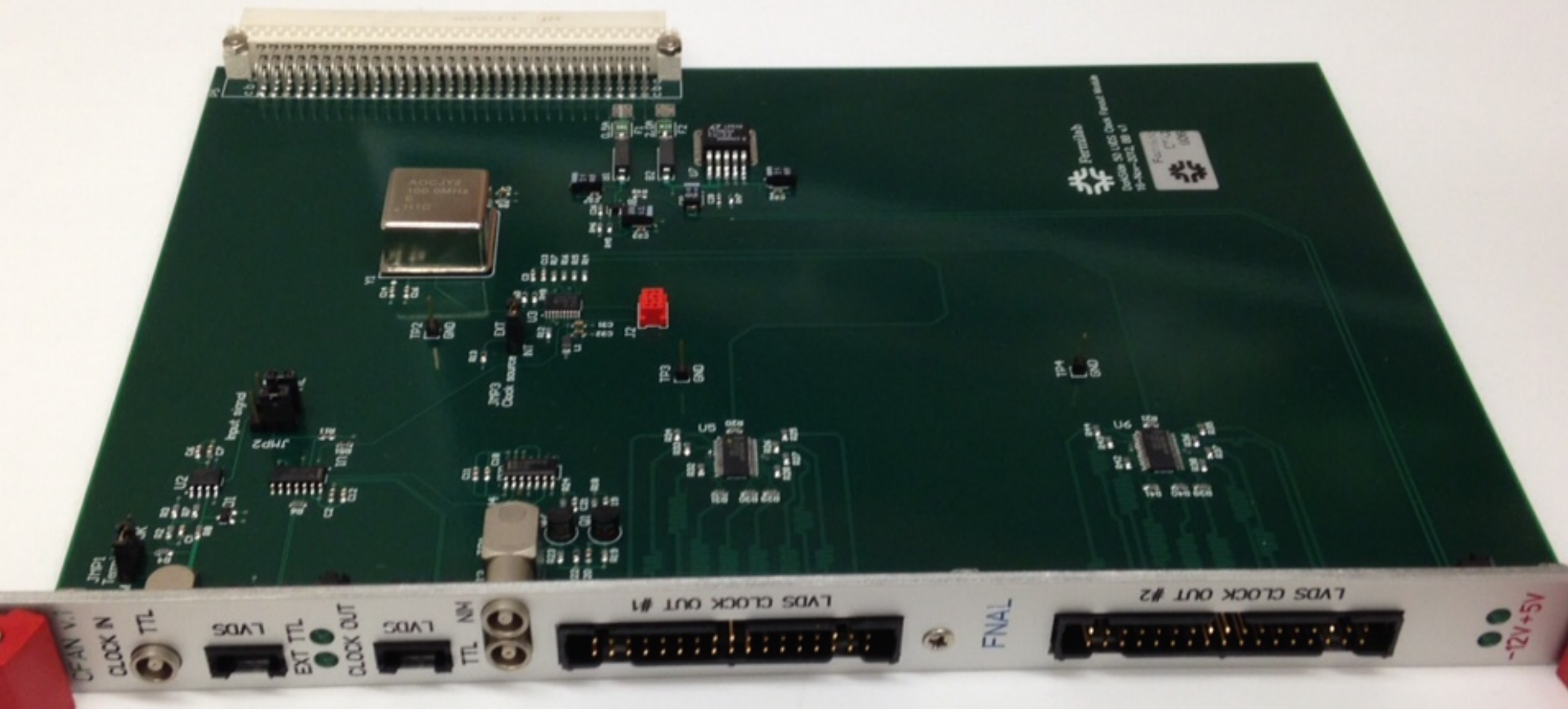}
\caption{The Clock Fanout Module}
\label{fig:TPCClockModuleSchematic-and-Photo}
\end{figure}

The final custom module in the TPC electronics system is the clock module used to synchronize the commercial digitizing modules and ensure that they sample at the identical rate.   The module is realized in VME 6 and produces a total of 20 LVDS outputs. It features a programmable clock synthesizer \cite{ref:ClockPLL}  and can use an external source or the on-board quartz crystal oscillator \cite{ref:Oscillator}. In practice, we use the internal source.  Figure \ref{fig:TPCClockModuleSchematic-and-Photo} shows the schematic and a photograph of the module.

\section{TPC Trigger}
\label{sec:TPC-Trigger}

The TPC trigger is based on the CAEN V1495 general purpose VME FPGA module ~\cite{ref:CAENV1495}. The module is programmed to provide:
\begin{itemize}
\item  a Majority Logic Trigger using the LVDS signals from the `front-end' modules as inputs.
\item  a Trigger Output signal used to initiate event recording. This signal can be derived from a number of trigger sources.
\item  Run Enable and Run Pause signals to control event recording.
\item  a Trigger Inhibit signal (that acts internally) to prevent further triggers when the digitizer memories are full 
\item  a 16-bit Trigger ID word which is included in the digitizer event header.
\item  three internal frequency generators for testing and calibration purposes 
\item  an internal 4k $\times$ 32 bit FIFO memory to store event meta-data.
\item a 1PPS timestamp for event synchronization
\item live time and dead time counters for live time corrections
\item a Trigger ID for the Veto system electronics \cite{ref:Vetoelectronics}
\item a 16-bit trigger pattern for testing the majority trigger logic. 
\item a `G2'  trigger based on the multiplicity of hits (LVDS signals from Front-end modules) within a fixed time window with two settable thresholds and two independent pre-scale values.
\end{itemize}

\begin{figure}[h]
\begin{center}
\includegraphics[width=0.7\textwidth]{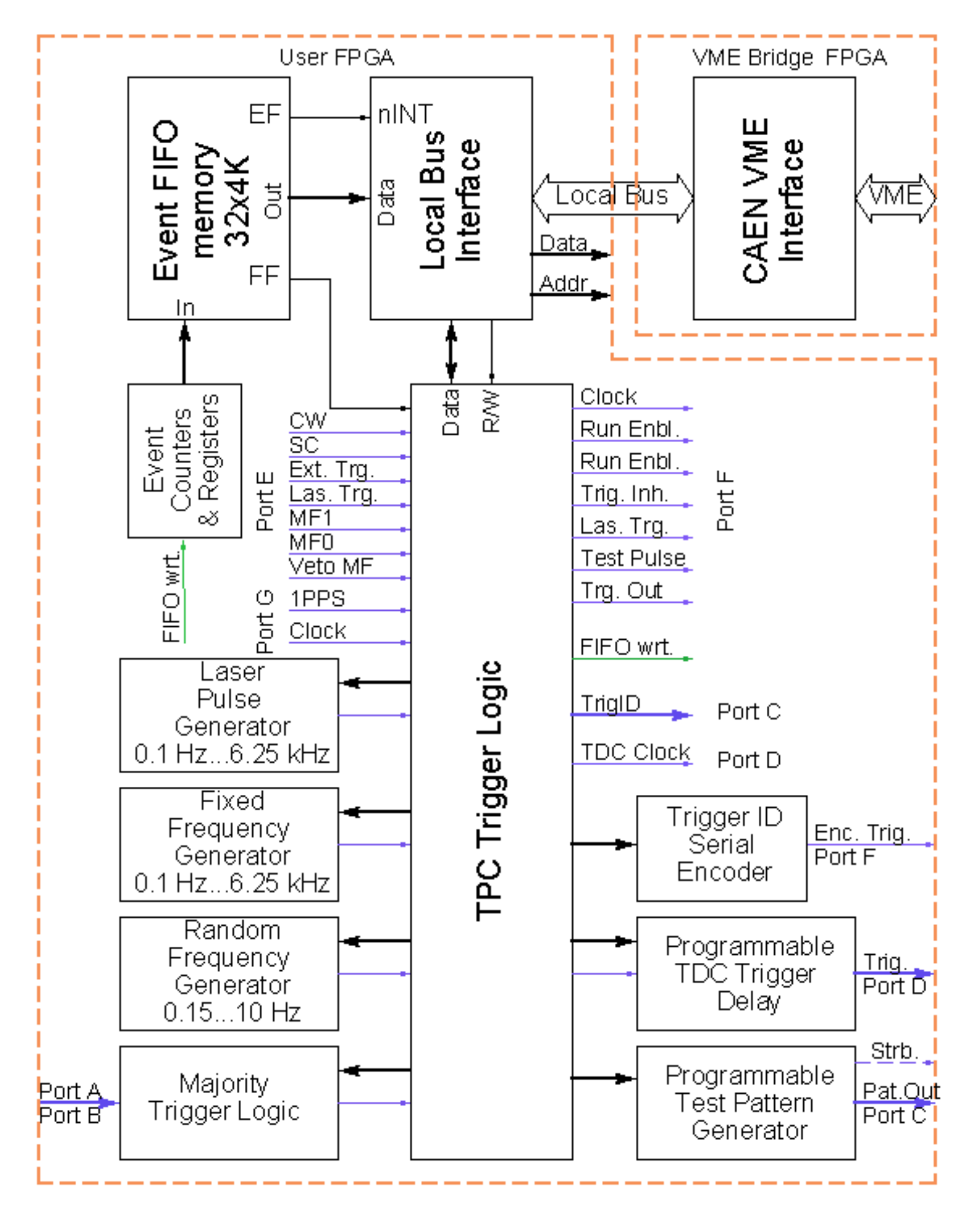}
\caption{TPC V1495 functions and schematic. The fine lines are signals; the broad lines are data words}
\label{fig:TPCV1495}
\end{center}
\end{figure}

A schematic representation of this functionality is given in Figure \ref{fig:TPCV1495}. 

The Majority Logic Trigger requires a user-settable (minimum) number of (different) channels to produce a signal within a user-settable time-window; the number of channels required can range from 1 to 38 (in unit steps) and the time window can range from 10 ns to 2400 ns in 10 ns steps.  Data have been taken with the number of channels required set to two and to three, and the time-window has been typically set to 100 ns.  
Table \ref{tab:V1495Majoritylogic} gives the inputs (software and hardware ) for the Majority Logic Trigger.

\begin{table}[h]
\begin{center}
\small
\begin{tabular}{| lll |}
\hline
Parameter & Value & Type \\  
 \hline
Inputs              		  		&  40 	 		& LVDS from Front-end module \\
Memory Full             		  	&  2 	 			& LVDS from TPC Digitizers and Veto system \\
Minimum Pulse width 		&  5 ns  			& LVDS output          \\
Channel Enable Mask	        	& 40 bits		 	& Software settable	\\
Majority Time Window		& 10 ns to 2550 ns	& Software Settable in 10 ns increments 	  \\
Majority Threshold			& 1 to 38			& Software settable \\
\hline\end{tabular}
\caption{V1495 inputs}
\label{tab:V1495Majoritylogic}
\end{center}
\end{table}

\begin{figure}[h]
\begin{center}
\includegraphics[width=0.97\textwidth]{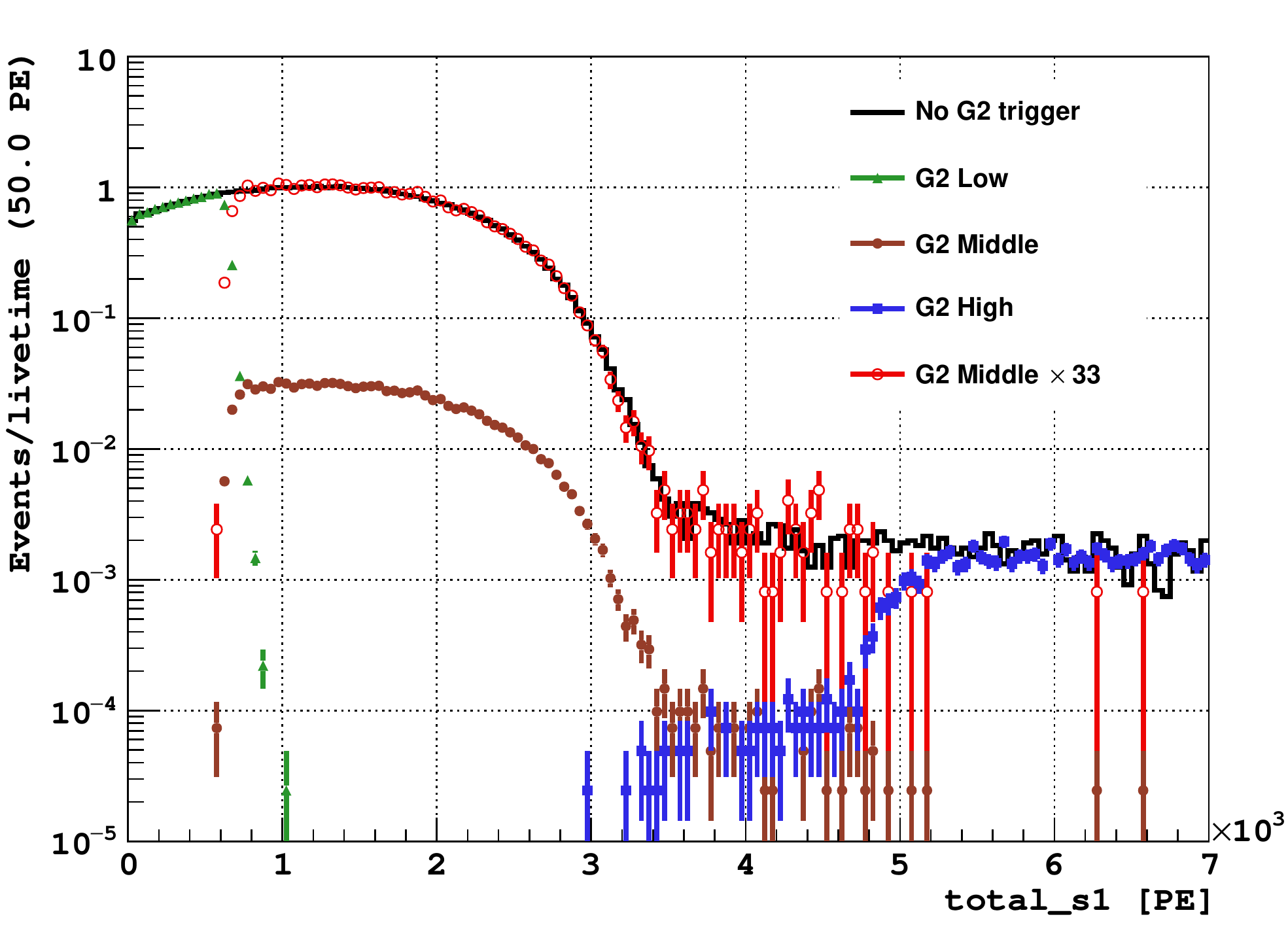}
\caption{Plot showing the effect of the G2 trigger set so that events between the lower threshold and the higher threshold were pre-scaled by 33. Black shows data taken with just the Majority Logic Trigger requirement before the G2 trigger was developed; the other colors show data taken with the G2 trigger; green shows  events identified as below the G2 lower threshold; brown shows data identified as above the G2 low threshold and below the G2 high threshold; blue shows events identified as above the G2 higher threshold; red shows the result of multiplying the prescaled data by the prescale factor}
\label{fig:TPCG2-trigger}
\end{center}
\end{figure}

The `G2' trigger mentioned serves as an example of the flexibility that the V1495 module affords. The trigger was developed for the running with atmospheric argon where $^{39}$Ar decays produced a rate of $\sim$45 Hz in the Majority Logic Trigger,  a rate that was not long-term sustainable in terms of data storage. The G2 trigger exploits the fact that the 
energy spectrum of  $^{39}$Ar decays extends a factor of $\sim6$ beyond the dark matter search energy region.  It uses the Majority Logic Trigger as pre-trigger and then counts the number of hits from all 38 PMT channels in a software settable time window.  The number of hits in this window is compared to two threshold values, and if the number of hits exceeds either threshold, prescale factors for each threshold are applied to decide if the event should be recorded. Five microseconds allows for a reasonable estimate of the light from the slow component of the argon scintillation and typical values for the thresholds have been 360 and 1500 hits with prescales of 33 and 1 respectively. This trigger effectively removed events from the top three-quarters of the  $^{39}$Ar decay energy spectrum, kept all the very high energy events,  and reduced the recording rate from 45 Hz by a factor of three without losing any efficiency in the dark matter search.  
Figure \ref{fig:TPCG2-trigger} shows the operation of the `G2' trigger. As well as its use with atmospheric argon, the G2 trigger has been exploited in calibrations with gamma sources (e.g. $^{22}$Na) where a cut in the energy spectrum has avoided recording data outside our region of interest.

For the running with low-radioactivity argon, which constitutes the majority of the dark matter search exposure, the rate due to $^{39}Ar$  is negligible \cite{ref:70days}. The event rate in this case is $\sim$ 1.5 Hz, from radioactivity in the detector materials, and the Majority Logic Trigger is quite adequate.

\section{DAQ}
\label{sec:DAQ}

The TPC data acquisition system is built to accommodate the 8-channel CAEN series V172x digitizers and their readout using the 4-channel CAEN  A3818C optical link and controller PCI E module \cite{ref:CAENA3818}. Two other modules, the CAEN V1190 TDC and the CAEN V1495 logic board, are readout through the CAEN 2718 VME Controller which itself is connected via optical link.  
There are a total of 10 digitizers, 5 @ 250 MHz 12 bit V1720 modules (range 2 Volts) \cite{ref:CAENV1720} and  5 @  100 MHz 14 bit V1724 modules (range 2.25 Volts) \cite{ref:CAENV1724}. They receive the high gain (100 MHz band width) and  the low gain (40 MHz bandwidth) analog signals  respectively from the front-end-modules. The discriminator outputs from the veto system \cite{ref:Vetopaper} are also recorded using two CAEN V1190 TDCs.  The event metadata (trigger type, trigger ordinal, live time since previous event, time since start of run, etc.) are generated and readout from the V1495.

\begin{figure}[h]
\begin{center}
\includegraphics[width=0.93\textwidth]{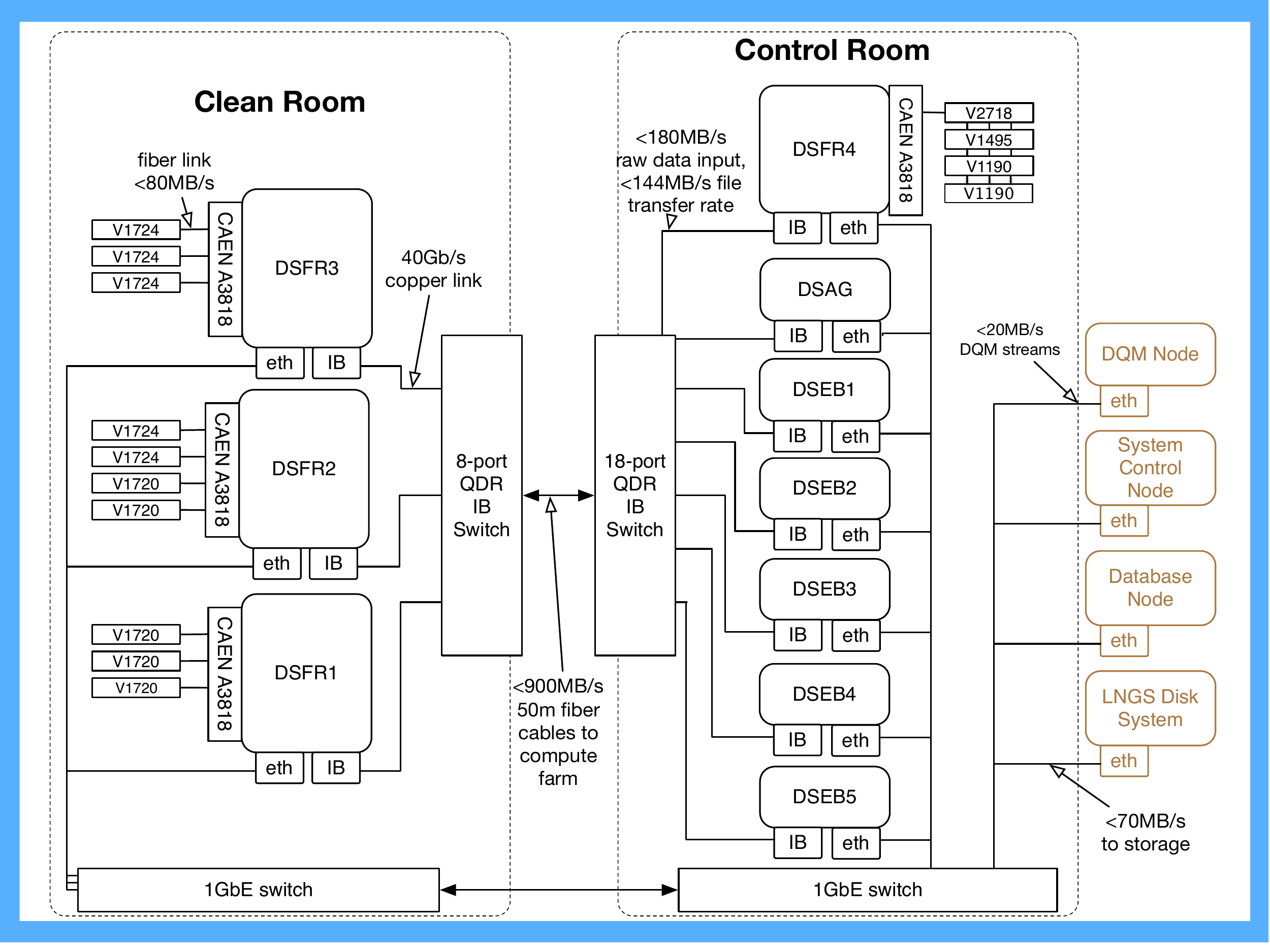}
\caption{The arrangement of computers in the TPC data acquisition system. The DSFRn are `Fragment Receivers' which read out the attached set of digitizers; the DSEBn are `Event Builders' which combine the fragments into complete event records. The DSAG computer collects the complete events, arranges them in real-time order and writes them to disk. Data are transferred from the FR computers to the EB computers via Infiniband - see text. }
\label{fig:computing_diagram}
\end{center}
\end{figure}

A schematic of the organization of the hardware is shown in Figure \ref{fig:computing_diagram}.

The readout system is highly parallel in order to achieve the maximum throughput.   The four `fragment receiver' computers, (the DSFRn in Figure \ref{fig:computing_diagram}), are each equipped with one 4 channel CAEN A3818C optical receiver.  The digitizer modules and the 2718 controllers are connected with their own individual optical link into an individual input of the A3818C. Within the fragment receiver computers, one process on one core is dedicated to one module (up to 4 dedicated cores in a single computer) and a fraction of another core is used for sending the data and internal monitoring. A set of high-speed links based on the Infiniband protocol \cite{ref:Infiniband} carries the event `fragments', essentially one module's data, from the fragment readers and distributes them to a number of event `builder' processes running on a second set of computers, the event builders (the DSEBn). 

The computers labeled DSFR1 to DSFR3 in Figure \ref{fig:computing_diagram}  are located in the clean room immediately above the water tank next to the TPC electronics. The computer DSFR4 is used to include a summary of data from the veto system into the TPC system data record and is located in the veto system electronics room.  The event builder computers (DSEBn) and the aggregator computer (DSAG) are also located in the veto electronics room

The event fragments are sent to the event builders in a simple round-robin scheme. The fragment sizes and number of event fragments per event are fixed for any given running mode and there is no arbitration scheme. The Infiniband link is well adapted to the MPI (Message Passing Interface) protocol and its use comes from the experience of developing computer systems at Fermilab for Lattice Gauge calculations \cite{ref:lattice-gauge}. Each event builder computer has four event building processes, each running on its own core. Every event fragment contains an event number and ID to remove any uncertainty about missing fragments in the event building, and the event building process takes less than a millisecond.   The complete (`built') events pass to the aggregator machine (DSAG) which puts the events in real-time order and writes them to disk. A second process running on the aggregator machine is responsible for the real-time monitoring. 

The data acquisition software is based on the Fermilab {\it{artdaq}} product\cite{ref:artdaq_writeup_CHEP}, a toolkit for creating data acquisition systems to be run on commodity servers.  It is written in C$++$ and provides reusable components for common DAQ functions and a framework for custom software components.

\begin{figure}[h]
\begin{center}
\includegraphics [width=0.95\textwidth]{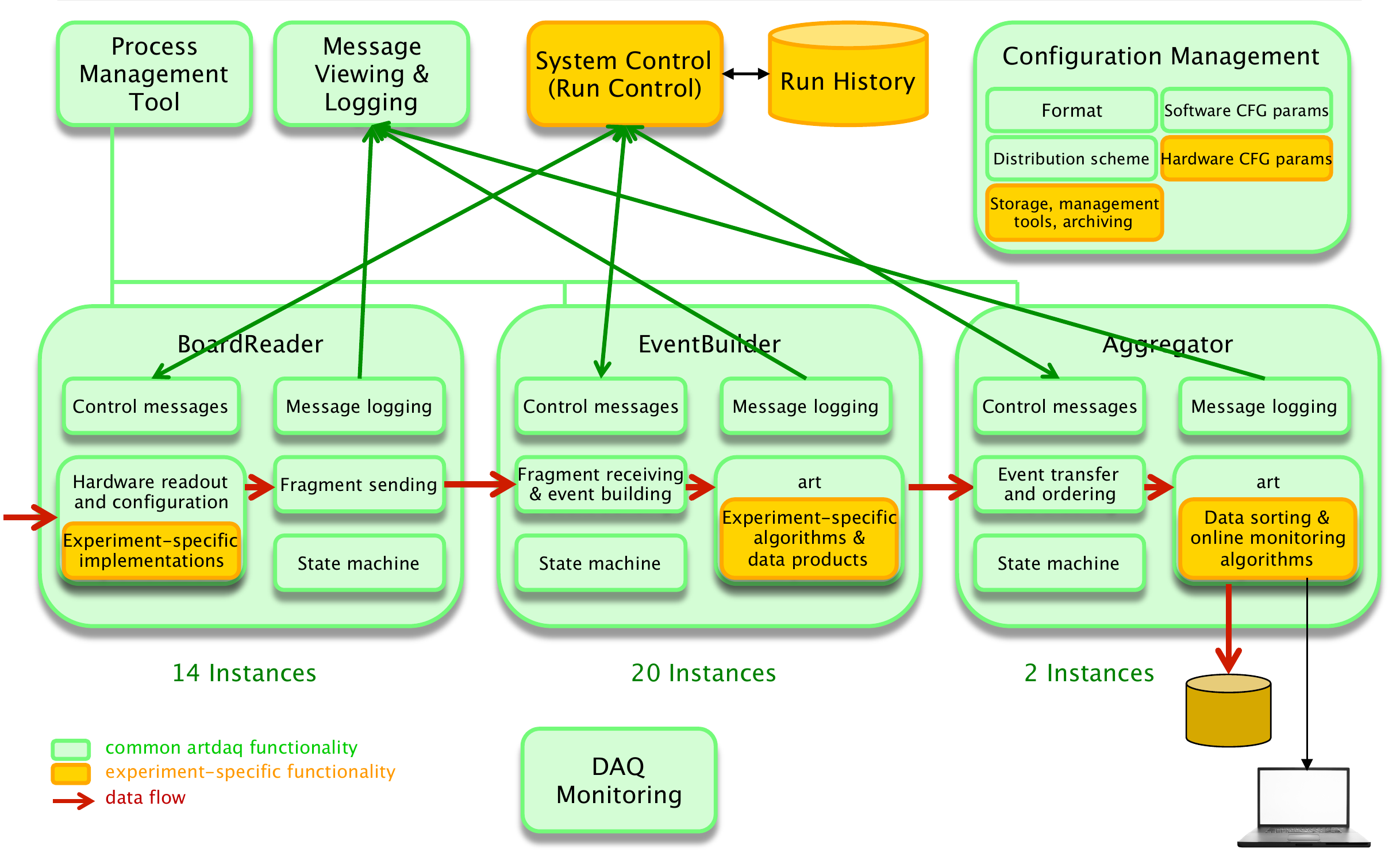}
\caption{Block diagram of the {\it{artdaq}} processes and software components.  Components that are shown in green are part of the base toolkit, ones shown in orange are developed for each experiment.  Data transfers are handled by an InfiniBand implementation of the Message Passing Interface (MPI).  Control messages are sent to individual {\it{artdaq}} processes using XMLRPC.}
\label{fig:artdaq}
\end{center}
\end{figure}

Figure \ref{fig:artdaq} shows a block diagram of the {\it{artdaq}} process types and the DAQ functions that they provide.   

BoardReader processes manage the configuration of the electronics and the readout of data fragments, and specific versions were developed for the V1720 and V1724 waveform ADCs, the V1190 TDC, and the V1495 programmable FPGA. The EventBuilder processes pass complete events to two instances of the {\it{art}} \cite{ref:art} analysis program, an analysis framework developed at Fermilab to allow for common use in online and offline environments. In our case,  {\it art} simply runs the Huffman compression algorithm and provides plots for the online monitoring displays.
 
The typical acquisition-gate width in the WIMP search mode, running the TPC at a drift voltage of 200V/cm, is 440 microseconds, giving a record length for each V1720 channel of 440$\times$250 samples = 220kB, and of 88 kB for the V1724, for a  total of 1760 kB and 704 kB respectively per 8-channel module. The raw event size is therefore  $5\times1760 + 5\times 704$ kB = 12.3 MB plus a small amount from the veto system summary and the metadata.  At the quoted 80MB/s of the CAEN optical readout protocol,  the V1720 digitizers transfer their data (in parallel) in about 20 milliseconds while the V1724 digitizers take about 8 milliseconds. (The burden of the V1495 and the V1190 readout is negligible.) The V1720 readout time therefore sets a maximum theoretical data rate of 48 events/s.  In practice readout rates of over 40 events/second have been achieved with little deadtime ($<20\%$) thanks to the large amount of memory in the V1720 digitizers and the power of the event-builder computers.  As with any buffered system, however, the livetime is high as long as the average input rate does not exceed the maximum output rate.  It should be noted, also, that while this generation of digitizers can perform pedestal suppression, this suppression is performed as the data are presented for output at the 80 MB/s rate and would serve to reduce the amount of data transmitted but does not reduce the time of the actual readout.

In our situation, however, a readout rate in excess of 500 MB/s, except in a short-term calibration mode, is not practical simply in terms of the subsequent data storage required. Despite our excellent signal to noise on a single-electron signal, we do not perform any pedestal suppression to ensure that we have the absolute maximum information about any given event. To reduce the data volume, a lossless Huffman compression is applied to each waveform.  The compression is tuned on a typical data set, and by applying it to the difference between successive digitizings in a given waveform, it needs to be calculated only once for each running condition even if the digitizer baseline changes. The compression reduces the data volume by a factor $\sim5$ resulting in an output event size of 2.5 MB.  When running with the atmospheric argon, we initially wrote data at more than 40 Hz accumulating, even with the compression, over 7 TB a day. The G2 trigger (see section~\ref{sec:TPC-Trigger}) allowed us to prescale events beyond our region of interest thereby reducing the rate of events acquired to 14 Hz and storing data at 2.5 TB/day.  As described, with the low-radioactvity argon in use for the major part of our WIMP search, the event rate is 1.5 Hz giving a data rate of 0.3 TB/day. These data are stored on disk at the LNGS central laborartory and sent to Fermilab where they are stored on tape.


\section{Summary}
\label{sec:Summary}

This article has described the general features of the electronics and data-acquisition used by the liquid argon time-projection chamber of the DarkSide experiment. The system includes a novel cryogenic pre-amplifier and operates with a signal to noise for a single photo-electron of better than 30 to 1 and a dynamic range of about 1500. The data-acquisition system can acquire 500 MB/s with a subsequent storage rate of 100 MB/sec after lossless compression, capabilities which have been exploited when the chamber was filled with atmospheric argon and for calibrations. The whole system has been operational since early 2014 and has proven reliable and robust since that time.

\newpage
\acknowledgments
The DarkSide-50 Collaboration  thanks the LNGS laboratory and its staff for their invaluable technical and logistical support. We particularly wish to thank our Fermilab colleagues Dr. Jin-Yuan Wu for his contributions to the FPGA firmware, Dr. Don Holmgren for his advice on the selection and procurement of the computing and networking hardware, Engineer Sten Hansen for his work on designing the PMT divider and his help with the V1495 firmware, and Engineer Ron Rechenmacher for resolving issues with Infiniband and our raid disk arrays. DarkSide-50 has been supported by the Italian Istituto Nazionale di Fisica Nucleare, the U.S. National Science Foundation under grant Nos. 1004051, 1242571, 1314268, 1314479, 1314483, 1314501, 1314507 \& 1314752,  the US Department of Energy under contracts DE-AC02-07CH11359 and DE-FG02-91ER40671, the Polish NCN (Grant UMO-2014/15/B/ST2/02561), and the Russian Science Foundation Grant No. 16-12-10369. We also acknowledge financial support from the UnivEarthS Labex program of Sorbonne Paris Cit\'e (ANR-10-LABX-0023 and ANR-11- IDEX-0005-02) and from the S\~ao Paulo Research Foundation (FAPESP) grant No. 2016/09084-0. Development of the in-liquid pre-amplifier was supported by the INFN CSN-V(5) QUPID-R\&D grant.

 \end{document}